\documentclass[journal,twoside,web]{ieeecolor}
\usepackage{generic}
\usepackage{cite}
\usepackage{amsmath,amssymb,amsfonts,comment,tabularx}
\usepackage{bbm}
\usepackage{bm}

\usepackage{url}
\usepackage{algorithmic}
\usepackage{graphicx}
\usepackage{textcomp}
\usepackage{xcolor}
\usepackage{hyperref}
\usepackage{caption}
\usepackage{subcaption}
\usepackage{stfloats}

\usepackage{multirow}

\usepackage{marginnote,zref-savepos}
\newcounter{labelnote}
\makeatletter
\let\oldmarginnote\marginnote
\renewcommand*{\marginnote}[1]{%
 \begingroup\strut
  \stepcounter{labelnote}\zsaveposx {marginnote-\thelabelnote}
     \ifnum 0\zposx{marginnote-\thelabelnote}<1900000
      \reversemarginpar
      \oldmarginnote{\color{blue}#1}%
     \else
      \normalmarginpar
      \oldmarginnote{\color{blue}#1}%
     \fi
 \endgroup%
}
\makeatother

\def\BibTeX{{\rm B\kern-.05em{\sc i\kern-.025em b}\kern-.08em
    T\kern-.1667em\lower.7ex\hbox{E}\kern-.125emX}}
\begin{document}
\title{On Exponential Utility and \\ Conditional Value-at-Risk as \\ Risk-Averse Performance Criteria}
\author{Kevin M. Smith and Margaret P. Chapman, \IEEEmembership{Member, IEEE} \vspace{-12mm}%
\thanks{This work was supported in part by the Computational Hydraulics International University Grant Program for complementary use of PCSWMM Professional software. K. M. Smith was supported by an NSF Integrative Graduate Education and Research Training award (NSF 0966093) and an NSF Research Traineeship award (NSF 2021874). The authors acknowledge the Tufts University High Performance Compute Cluster (\href{https://it.tufts.edu/high-performance-computing}{https://it.tufts.edu/high-performance-computing}) which was utilized for the research reported in this paper. The work of M. P. Chapman was supported in part by the Edward S. Rogers Sr. Department of Electrical and Computer Engineering, University of Toronto. M. P. Chapman acknowledges the Natural Sciences and Engineering Research Council of Canada Discovery Grants Program [RGPIN-2022-04140]. M. P. Chapman reconna\^{i}t le Conseil de Recherches en Sciences Naturelles et en G\'{e}nie du Canada.}
\thanks{K. M. Smith is with the Department of Civil and Environmental
Engineering, Tufts University, Medford, MA 02155 USA and OptiRTC, Inc., Boston, MA 02114 USA (email: kevin.smith@tufts.edu).}
\thanks{M. P. Chapman is with the Edward S. Rogers Sr. Department of Electrical and Computer Engineering, University of Toronto, Toronto, ON M5S 3G4 Canada (email: mchapman@ece.utoronto.ca).} 
}
\maketitle
\pagestyle{empty}
\thispagestyle{empty}
\begin{abstract}
The standard approach to risk-averse control is to use the \emph{Exponential Utility} (EU) functional, which has been studied for several decades. %
Like other risk-averse utility functionals, EU encodes risk aversion through an increasing convex mapping $\varphi$ of objective costs to subjective costs. An objective cost is a realization $y$ of a random variable $Y$. In contrast, a subjective cost is a realization $\varphi(y)$ of a random variable $\varphi(Y)$ that has been transformed to measure preferences about the outcomes. For EU, the transformation is $\varphi(y) = \exp(\frac{-\theta}{2}y)$, and under certain conditions, the quantity $\varphi^{-1}(E(\varphi(Y)))$ can be approximated by a linear combination of the mean and variance of $Y$. More recently, there has been growing interest in risk-averse control using the \emph{Conditional Value-at-Risk} (CVaR) functional. In contrast to the EU functional, the CVaR of a random variable $Y$ concerns a fraction of its possible realizations. If $Y$ is a continuous random variable with finite $E(|Y|)$, then the CVaR of $Y$ at level $\alpha$ is the expectation of $Y$ in the $\alpha \cdot 100 \%$ worst cases. Here, we study the applications of risk-averse functionals to controller synthesis and safety analysis through the development of numerical examples, with emphasis on EU and CVaR. Our contribution is to examine the decision-theoretic, mathematical, and computational trade-offs that arise when using EU and CVaR for optimal control and safety analysis.
We are hopeful that this work will advance the interpretability and elucidate the potential benefits of risk-averse control technology. %
\end{abstract}

\begin{IEEEkeywords}
Conditional Value-at-Risk, Exponential Utility, Risk aversion, Safety analysis, Stochastic systems.
\end{IEEEkeywords}
\newtheorem{algorithm}{Algorithm}
\newtheorem{problem}{Problem}
\newtheorem{theorem}{Theorem}
\newtheorem{remark}{Remark}
\newtheorem{lemma}{Lemma}
\newtheorem{assumption}{Assumption}
\newtheorem{proposition}{Proposition}
\newtheorem{definition}{Definition}
\section{Introduction}
\label{sec:introduction}

\IEEEPARstart{W}{hile} there is no universal definition of risk, there is growing recognition that measures of risk should be informed by both the \emph{probability} and \emph{severity} of harmful events.\footnote{This recognition is evident in the changing definitions of risk codified by the International Organization for Standardization. Compare the definition of risk as the ``probability of loss or injury from a hazard'' in \cite{ISO22538} to a more contemporary standard that accounts for ``consequences'' as well as their ``likelihood,'' e.g., see \cite{ISO31000}.} This represents a departure from traditional definitions of risk in many engineering disciplines, where risk has often been expressed in terms of probability alone, e.g., see \cite{read2015}. The expanded definition of risk raises important questions about how to combine the probability and severity of outcomes into a useful risk measure. One approach is to simply measure risk as the expected cost, that is, the probability-weighted average of the outcomes. Such a measure is said to be \emph{risk-neutral} because it is insensitive to the characteristics of the outcome distribution (e.g., spread, higher-order moments, etc.) around the expected cost. In contrast, we use the term \emph{risk-sensitive} to describe risk measures that are responsive to these characteristics.

Here, we are mainly concerned with \emph{risk aversion}, a type of risk sensitivity that generally prefers outcome distributions with smaller spreads and tail-costs (for the same expected cost). Incorporating risk aversion into the analysis and synthesis of control systems provides a potentially useful alternative to risk-neutral or worst-case methods. Indeed, summarizing a random outcome in terms of its expectation neglects other characteristics of its distribution that may have practical importance. For example, 
 average performance measures can mask the presence of rare outcomes that would be considered ruinous and unacceptable.
On the other hand, approaches that focus solely on the worst-case outcome may be too sensitive to the estimate of that outcome, or lead to designs that are unnecessarily conservative or too expensive to implement in practice. All together, the above limitations motivate the investigation of risk-averse methods that allow a decision-maker some flexibility between these two extremes.

Since the 1970s, algorithms have been developed to minimize a random cost incurred by a control system, where the cost is assessed in terms of a \emph{risk-averse functional} \cite{jacobson1973, whittle1990risk, howardmat1972, masi1999, ruszczynski2010risk, bauerleott, bauerlerieder, haskell, van2015distributionally, samuelson2018safety, borkar, chow2015risk, pflug2016timeEuro}. Recent work has also sought to incorporate insights from prospect theory into control systems by making them risk-seeking towards some outcomes while being risk-averse towards others \cite{lilianratliff}.\footnote{Risk-seeking control is the other form of risk-sensitive control. Under risk-seeking control, larger spreads in the outcomes are assumed to represent opportunities rather than liabilities, and tail-rewards are considered to be more important than tail-costs.} A survey of approaches to risk-sensitive control from an optimal control perspective can be found in our recent work \cite{wangchapman}. 

Nonetheless, the question of when a particular risk-averse functional may be more suitable for the analysis or synthesis of a control system is not well-understood. We seek to shed light on this question by focusing our study on optimal control using Exponential Utility and Conditional Value-at-Risk, which are arguably the two most popular and well-established risk-averse functionals.

\emph{Exponential Utility} (EU) is the classical risk-averse functional in control engineering. A value iteration algorithm, where the domain of the value functions is the state space of a control system, can be derived to solve an EU-optimal control problem, e.g., see \cite{whittle1990risk, masi1999, bauerlerieder, classicalrstheorypaper}. Like other utility functionals, EU maps objective costs to subjective costs. To encode ``rational'' risk aversion, risk-averse utility functionals transform objective costs using an increasing and convex mapping $\varphi$ 
\cite[Chap. 1]{eeckhoudt2005}. EU is of particular interest because under certain conditions, the quantity $\varphi^{-1}(E(\varphi(Y)))$ can be approximated by a linear combination of the mean and variance of a random cost $Y$. %
This provides a concise interpretation of risk-averse EU-optimal control as an approximation for a multi-objective mean-variance minimization problem. However, outside of the limited set of conditions where this interpretation is valid, EU-optimal control can only be said to minimize an infinite linear combination of moments. This latter case arguably makes EU harder to interpret, parametrize, and identify practical situations in which its use is appropriate.

In financial portfolio optimization, it is conventional to approximate the first and second moments of random returns of financial assets directly rather than compute an expected utility.
In this application, which typically involves the optimization of one decision at one time, approximating the first two moments may be preferred because choosing a utility function and knowing distributions exactly are not required, e.g., see \cite{mark2014} and the references therein. This perspective is called the \emph{Markowitz Model}, which is founded on the seminal work by H. Markowitz from 1952 \cite{Markowitz1952}. 

The optimal control problem of minimizing variance subject to an equality constraint on the mean can be solved efficiently in a linear-quadratic setting via Riccati equations \cite{won2001, costcumchapter}.
More generally, mean-variance optimal control is computationally expensive, as it cannot be solved using dynamic programming (DP) on the original state space. Miller and Yang, for instance, develop an interesting bilevel optimization approach \cite{milleryang}. The investigation of EU-optimal control has been motivated by its theoretical connections to mean-variance and its computational simplicity. %

EU-optimal control was first studied in the context of finite state spaces by Howard and Matheson in 1972 \cite{howardmat1972}. EU was applied to optimal control of linear systems with quadratic costs on continuous state spaces by Jacobson in 1973 \cite{jacobson1973}. This theory was extended by Whittle and colleagues in the 1980s and 1990s, in particular, to the setting of partially observable systems \cite{whittle1981, whittle1986, whittle1990risk, whittle1991}. In the context of linear systems with quadratic costs subject to Gaussian noise, EU-optimal control is often called LEQR or LEQG. The minimum entropy $\mathcal{H}_\infty$ controller and the infinite-time LEQR controller are equivalent \cite{gloverdoyle1988, Mustafa1990}. Connections between minimax model predictive control and model predictive control with an EU objective have been studied for linear systems with quadratic costs \cite[Chap. 8.3]{minimaxmpc}. 
EU-optimal control is a special case of mixed $\mathcal{H}_2$/$\mathcal{H}_\infty$ control synthesis in the linear-quadratic case (see \cite{zhang2019} and the references therein), cost-cumulant control \cite{Pham2002}, \cite{costcumchapter}, \cite{Cosenza2008}, and optimizing an expected utility \cite{bauerlerieder}. The EU functional has been applied to, for example, missile guidance \cite{Speyer1976}, inventory control \cite{Godoy1998}, control of active suspensions on vehicles \cite{Brezas2013}, and control of satellite attitude \cite[Sec. 5.6.1]{costcumchapter}. 

While expected utility optimization dates back to the 1950s, if not earlier, the optimization of \emph{Conditional Value-at-Risk} (CVaR) was not studied until the early 2000s \cite{Rockafellar2000, rockafellar2002conditional, acerbitasche}. This functional has been studied primarily by the operations research and financial engineering communities. The CVaR of a random cost represents the expected cost in a given fraction of the worst outcomes. Unlike EU-optimal control, CVaR-optimal control problems do not satisfy Bellman's Principle of Optimality on the original state space in general \cite{bodafilar}; the term \emph{time-inconsistent} is used to describe such problems \cite{shapirotime}. One approach for overcoming this issue is to utilize state-space augmentation.\footnote{While we focus on discrete-time systems with stage and terminal costs, we note that CVaR-optimal control problems for continuous-time systems with terminal costs have been studied  without state-space augmentation \cite{milleryang}.} This approach involves specifying the dynamics of a system on an enlarged state space so that Bellman's Principle is, in fact, satisfied for value functions that are defined appropriately on the augmented space. One can guarantee the existence of an optimal policy that depends on the augmented state dynamics under a measurable selection condition \cite{bauerleott, bauerlerieder, bauspectral}; such a policy may be called an optimal pre-commitment policy to emphasize its extra dependencies. A state-augmentation approach has been used to solve a CVaR-optimal control problem exactly in \cite{bauerleott} and approximately in \cite{chow2015risk}.
A related approach for solving a CVaR-optimal control problem (on an infinite time horizon) is to pose an infinite-dimensional linear program in occupation measures on an augmented state space \cite{haskell}. %

Another line of research has focused on minimizing an expected cumulative cost incurred by a stochastic system subject to a CVaR constraint, e.g., see \cite{borkar,van2015distributionally,samuelson2018safety}. In particular, Samuelson and Yang used this formulation to define a stage-wise safety specification, where the CVaR of the stage cost at time $t$ must be sufficiently small for each $t$ \cite{samuelson2018safety}.

More broadly, research on safety analysis for control systems has been active since at least the 1970s. Bertsekas and Rhodes proposed a safety analysis method for discrete-time systems using robust (minimax) optimal control in 1971 \cite{bertsekas1971minimax}. A safety analysis method for continuous-time systems with bounded disturbances via Hamilton-Jacobi (HJ) equations was introduced in the mid-2000s \cite{mitchell2005time}. This method, which is called HJ reachability analysis, has been further developed theoretically and in applications over the last decade, e.g., see \cite{lygeros2011, chen2018decomp, chen2018hamilton} and the references therein. %
The above two methods allow the computation of a set of initial conditions from which an uncertain system reaches a target set or avoids an unsafe region when subject to bounded adversarial disturbances. Generally, these disturbances lack probabilistic descriptions and are assumed to realize their most detrimental values.

Abate et al. proposed a less conservative safety analysis method for stochastic systems in 2008 \cite{abate2008probabilistic}. Stochastic safety analysis allows one to compute a set of initial conditions from which a system's probability of avoiding an unsafe region is sufficiently large \cite{abate2008probabilistic}. This method has been extended to reach-avoid and distributionally robust settings \cite{summers2010verification, ding2013, yang2018dynamic}. 

Using risk-averse functionals to define safety specifications for control systems is a relatively new idea and is motivated by the practical importance of quantifying both the probability and severity of harmful outcomes. Examples from the literature include \cite{samuelson2018safety} and our prior work \cite{chapmanACC, chapmantac2021, chapmantac2022}. Ref. \cite{samuelson2018safety} proposed a stage-wise, risk-averse safety specification (mentioned previously), whereas we proposed a trajectory-wise, risk-averse safety specification \cite{chapmanACC, chapmantac2021, chapmantac2022}. In this prior work, we considered the problem of minimizing the CVaR of a maximum cost of the state trajectory. We defined risk-sensitive safe sets as level sets of the optimal value function, and we derived methods for their estimation \cite{chapmanACC, chapmantac2021} and computation \cite{chapmantac2022}. %

Here, our contribution is to examine the decision-theoretic, mathematical, and computational trade-offs that arise when using EU and CVaR for optimal control and safety analysis. We illustrate such trade-offs by developing numerical examples of a thermostatic regulator and a stormwater system with a cumulative cost. In particular, we study how the empirical statistics of an optimal cost distribution (for a given control system and risk-averse functional) vary as the level of risk aversion varies.\footnote{The use of a Pareto frontier to assess trade-offs between competing objectives is common, for example, in reservoir management and financial portfolio optimization \cite{Castelletti, Lotov1998, Lotov2005, mark2014, Markowitz1952}.} 
We investigate the degree to which risk-averse EU-optimal control provides a useful approximation to mean-variance multi-objective optimization.

\emph{Notation}. If $S$ is a metrizable space, $\mathcal{B}_S$ is the Borel sigma algebra on $S$.\footnote{$\mathbb{R}^n$ with the Euclidean metric, and more generally any metric space, is a metrizable space. Informally, $\mathcal{B}_S$ is a large collection of subsets of $S$ that are ``regular enough'' to be measured. For a formal definition of $\mathcal{B}_S$ and further details about measure-theoretic concepts, please refer to \cite{bertsekas2004stochastic}, for example.} $\mathbb{R}_+^n := \{ x \in \mathbb{R}^n : x_i \geq 0, \; i = 1,2,\dots,n \}$ is the non-negative orthant in $\mathbb{R}^n$. We define $\mathbb{T} := \{0,1,\dots,N-1\}$ and $\mathbb{T}' := \{0,1,\dots,N\}$, where $N \in \mathbb{N}$ is given. We use the abbreviations: w.r.t. = with respect to, s.t. = such that, a.e. = almost everywhere or almost every, l.s.c. = lower semi-continuous, and cfs = cubic feet per second.

\emph{Organization}. Sec. \ref{SecII} studies how EU and CVaR encode risk aversion, and Sec. \ref{SecIII} presents algorithms for EU- and CVaR-optimal control. Sec. \ref{sysmodels} provides models of a thermostatic regulator and a stormwater system. Sec. \ref{numresults_optimal_control} develops numerical examples of optimal control. Sec. \ref{SecIV} focuses on safety analysis, and we provide concluding remarks in Sec. \ref{SecV}. 

\section{Quantification of Risk Aversion}\label{SecII}
Consider a random variable $Z$, representing a cost, that arises as a control system operates over time. $Z$ representing a cost means that smaller realizations of $Z$ correspond to better outcomes in the real world, whereas larger realizations of $Z$ correspond to worse outcomes. A standard stochastic control problem is to minimize the expectation of $Z$, subject to a given dynamics model, over a class of control policies (e.g., deterministic Markov). Different distributions can have the same expectation, but such distributions appear equivalent in the context of this problem. Focusing solely on the expectation of $Z$ ignores other characteristics of $Z$ (e.g., spread, higher-order moments) that may have practical significance. In this sense, minimizing the expectation of $Z$ is considered to be \emph{risk neutral}. In contrast, we use the term \emph{risk sensitive} to describe control problems that are aware of these characteristics. Here, we are mainly concerned with \emph{risk-averse control}, a form of risk-sensitive control that penalizes outcome distributions with larger spreads and tail-costs (for the same expected cost). Specifically, we focus on formulating risk-averse control problems using the Exponential Utility (EU) and  Conditional Value-at-Risk (CVaR) functionals.

\begin{figure}[b!]
\includegraphics[trim={0cm 2cm 1cm 0.5cm},clip,width=\columnwidth] {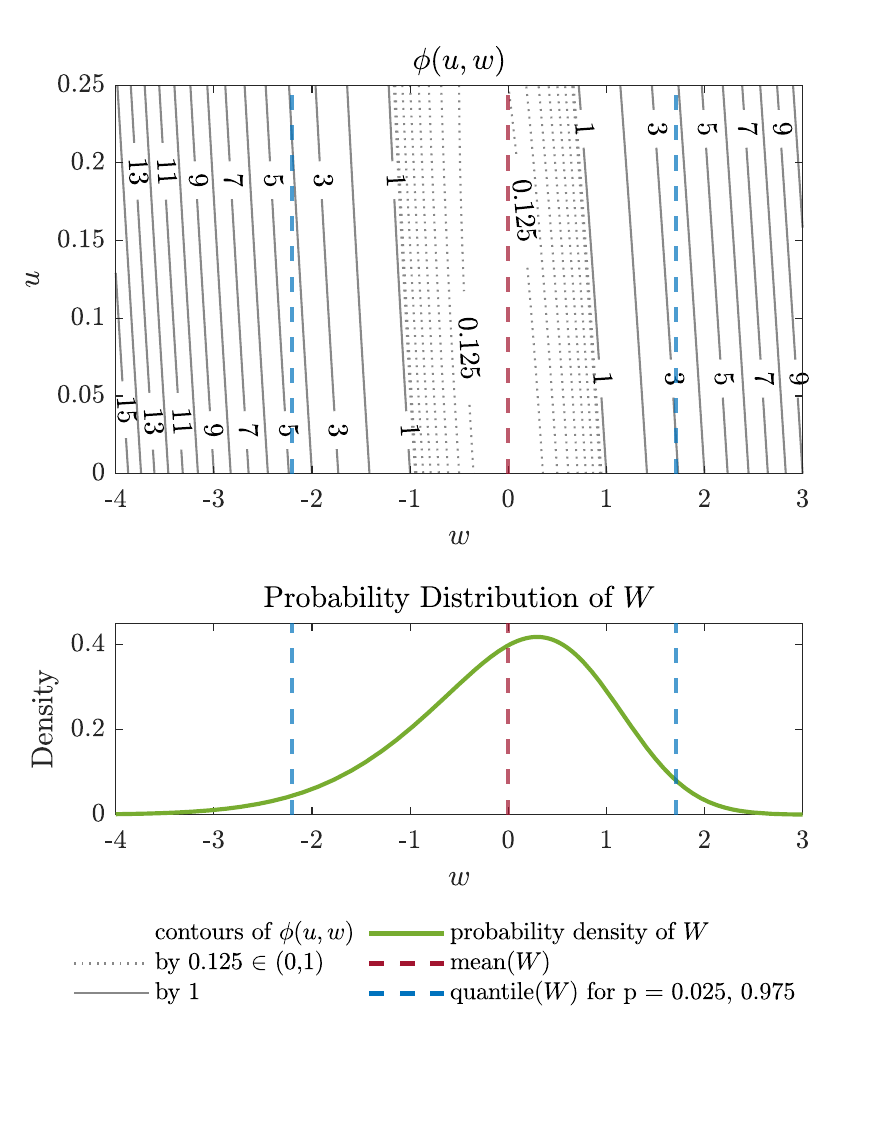}
\caption{\textbf{Top:} A contour plot of $\phi$ with vertical reference lines drawn at the 2.5\%, 50\%, 97.5\% quantiles of $W$. Note how the values of $\phi(u,W)$ change along these reference lines. The value of $\phi(u,W)$ increases slightly with increasing $u$ when $W$ takes on its mean value of 0. However the value of $\phi(u,W)$ decreases (increases) more dramatically with increasing $u$ at the 2.5\% (97.5\%) quantiles of $W$. \textbf{Bottom:} A probability plot of $W$ with the same quantiles marked.}
\label{skewnormal_contours_example}
\end{figure}

\subsection{Pedagogical Example}
Before presenting these functionals formally, we provide an example. Suppose that we would like to choose an input $u \in \mathbb{R}$ to minimize a quadratic cost, $\phi(u,w) := u^2 + (w + u)^2$, where $w \in \mathbb{R}$ is an unknown value of a random disturbance $W$ with zero mean. 
If our preferences are risk neutral, the optimal choice for minimizing the expectation of $\phi(u,W)$ is $u = 0$, regardless of the characteristics of $W$ other than its mean. For example, assume that $W$ follows a zero-mean skew normal distribution with unit variance and a skewness of $-0.5$. This distribution is shown in Fig. \ref{skewnormal_contours_example}, below a contour plot of $\phi$. The intersections of the vertical reference lines with the $\phi$-contours show how the value of $\phi(u,W)$ changes with $u$ and how this change depends on the quantile of $W$. From a risk-neutral perspective, the effect of $u$ on the quantiles of $\phi(u,W)$ is not important because only the expectation of $\phi(u,W)$ is of interest, and the expectation is minimized when $u=0$.

\begin{figure}[t]
\centerline{\includegraphics[trim={0.35cm 0 1cm 0},clip,width=\columnwidth]{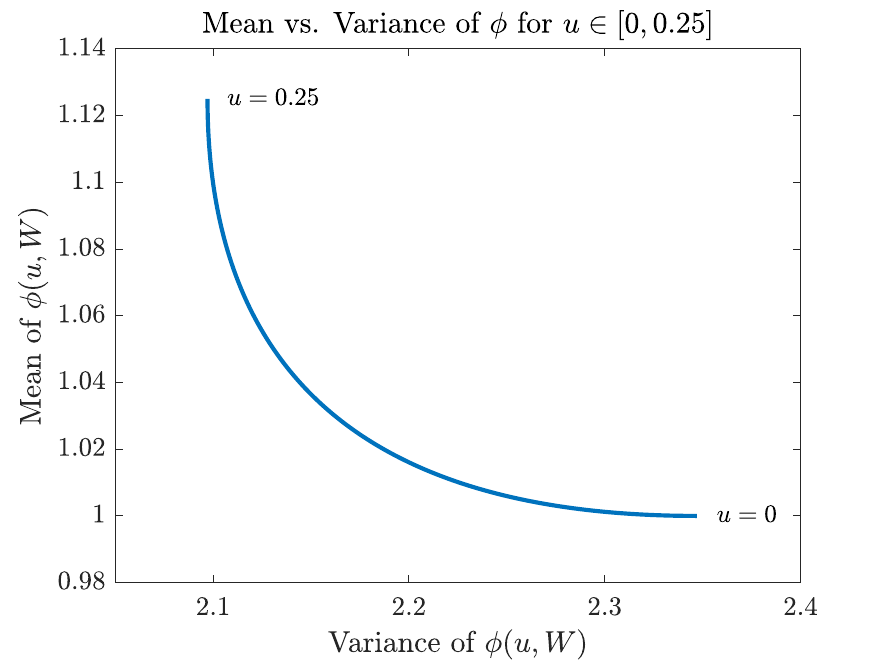}}
\caption{This plot presents the trade-off between the mean and variance of $\phi(u,W)$ for values of $u$ in the range [0, 0.25].} 
\label{skewnormal_mean_variance_tradeoff}
\end{figure}

\begin{figure}[t]
\centerline{\includegraphics[trim={.6cm 0 0.8cm 0},clip,width=\columnwidth]{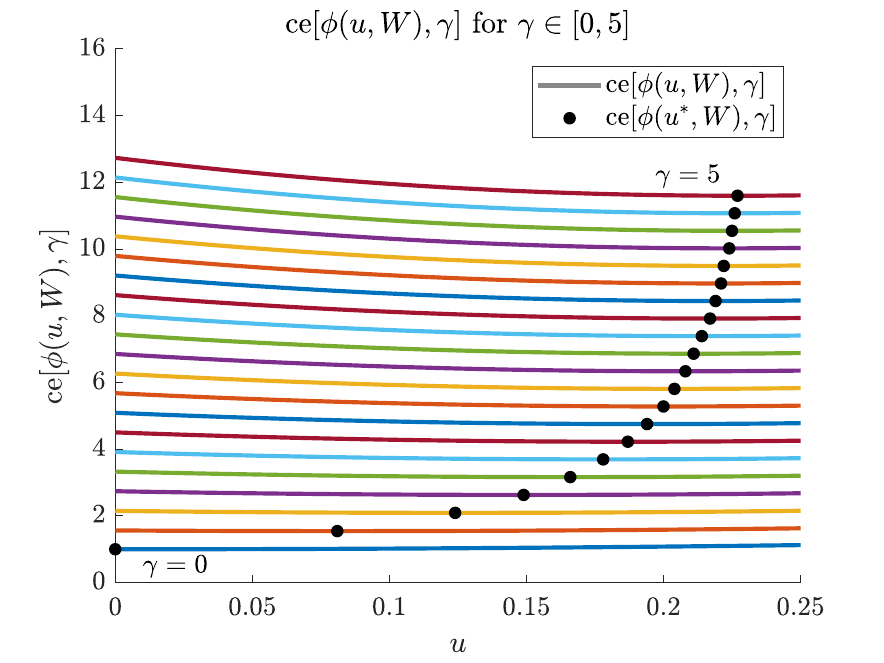}}
\caption{These curves show the certainty equivalent values of $\phi(u,W)$ for different values of $\gamma$. The certainty equivalence relation is defined in \eqref{define_ce}, and $\gamma$ reflects the price one is willing to pay to reduce the variance of $\phi(u,W)$.}
\label{skewnormal_ce_curves}
\end{figure}

However, from a risk-averse perspective, this effect is worth investigating, as there may be potential to reduce undesirable features (e.g., variance, average tail risk, etc.) of the distribution of $\phi$ at the expense of increases in the expectation. For example, if $u \in [0, 0.25]$, then there is a trade-off where larger values of $u$ can reduce the variance of $\phi(u,W)$ at the expense of increasing its expected value (Fig. \ref{skewnormal_mean_variance_tradeoff}). The amount of increase in the expectation that we are willing to endure for a reduction in the variance depends on our \emph{risk preferences}. For example, there may be a constant price $\gamma$ that we are willing to pay. Then, we can define a \emph{certainty equivalent} objective function to minimize:
\begin{equation}\label{define_ce}
    \text{ce}_\gamma(\phi(u,W)) := E(\phi(u,W)) + \gamma \cdot \text{var}(\phi(u,W)).
\end{equation}
The above relation describes a risk-averse actor's indifference between incurring a certain cost $\text{ce}_\gamma(\phi(u,W))$ or an uncertain cost $\phi(u,W)$. The left hand side of \eqref{define_ce} is called the \emph{certainty equivalent}. This value reflects the maximum price a rational risk-averse actor is willing to pay for insurance to avoid an uncertain outcome. In this sense, the certainty equivalent is a measure of the perceived risk of $\phi(u,W)$, and a risk-averse actor may choose a value $u$ to minimize this risk. Fig. \ref{skewnormal_ce_curves} shows plots of \eqref{define_ce} as a function of $u$ for a fixed $\gamma \in [0, 5]$ and identifies the optimal values of $u$ that minimize the certainty equivalent $\text{ce}_\gamma(\phi(u,W))$. %

\begin{figure}[!t]
\centerline{\includegraphics[trim={0.8cm 0 0cm 0},clip,width=\columnwidth]{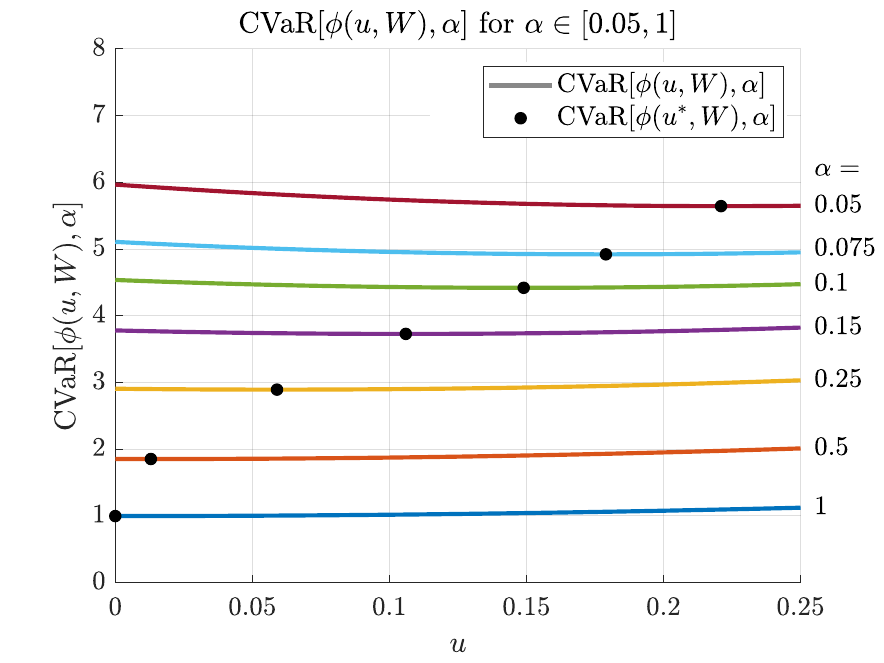}}
\caption{These curves depict the CVaR of $\phi(u,W)$ for different values of $\alpha$. CVaR is given by \eqref{define_cvar_informal}, and $\alpha$ reflects the fraction of the worst outcomes considered.}
\label{skewnormal_cvar_curves}
\end{figure}

However, risk-averse actors can perceive risk in features of $\phi(u,W)$ other than its mean and variance. For example, a risk-averse actor may express risk in terms of expected costs in the upper tail of $\phi(u,W)$. One such measure is Conditional Value-at-Risk. Let $F$ be the cumulative distribution function (CDF) of $\phi(u,W)$ and $F^{-1}$ its generalized inverse, the quantile function of $\phi(u,W)$. ($F$ and $F^{-1}$ depend on $u$ and $W$, which we do not write for brevity.) Then, the CVaR of $\phi(u,W)$ at level $\alpha \in (0,1]$ is given by
\begin{equation}\label{define_cvar_informal}
\text{CVaR}_{\alpha}(\phi(u,W)) = \textstyle \frac{1}{\alpha} \int_{1-\alpha}^{1} F^{-1}(\ell) \; \mathrm{d}\ell.
\end{equation}
CVaR encodes risk by assessing the expected costs in the worst $\alpha \cdot 100\%$ of values of $\phi(u,W)$.
By choosing to minimize CVaR for a particular $\alpha$, one expresses a desire to minimize the expected cost in one fraction of the worst values, even if doing so increases the expected costs in other fractions. In this way, $\alpha$ captures a preference for a desired level of risk aversion.

 Let $u \in [0, 0.25]$ be given. As Fig. \ref{skewnormal_cvar_curves} demonstrates, one can adopt a risk-neutral perspective and minimize the expected cost over all possible outcomes of $\phi(u,W)$ (i.e., $\alpha = 1$) by selecting $u = 0$. Alternatively, one can adopt a risk-averse perspective and focus only on minimizing the expected cost in a smaller fraction ($\alpha < 1$) of the worst outcomes by selecting a larger value of $u$. It is important to note that the choice of a risk-aversion level often involves a trade-off. In this example, any $u$ that minimizes CVaR for a particular choice of $\alpha$ leads to a non-optimal CVaR at all other levels of $\alpha$.

This example demonstrates two distinct ways in which risk preferences can be incorporated into an objective function: 1) through a weighted sum of expectation and variance or 2) through an average of a fraction of largest costs. In certain circumstances, the first approach corresponds to minimizing Exponential Utility, while the second approach corresponds to minimizing CVaR. As we explore in this paper, EU- and CVaR-optimal control approaches offer significant differences in their interpretatibility and computational efficiency when applied to more complicated examples.

\subsection{Control System Model}\label{controlsysmodel}
Now, we consider the case in which $Z$ is a random variable, representing a cost, that arises as a control system operates over time. In particular, we consider a system on a discrete, finite time horizon of length $N \in \mathbb{N}$ of the form
    $x_{t+1} = f(x_t, u_t, w_t)$ for $t = 0,1,\dots,N-1$,
where $x_t \in S$, $u_t \in A$, and $w_t \in D$ are realizations (i.e., values) of the random state $X_t$, the random control $U_t$, and the random disturbance $W_t$, respectively. The state space $S$, the control space $A$, and the disturbance space $D$ are Borel spaces; e.g., $\mathbb{R}^n$ and $B \in \mathcal{B}_{\mathbb{R}^n}$ are Borel spaces \cite[Def. 7.7, p. 118, Prop. 7.12, p. 119]{bertsekas2004stochastic}. The initial state $X_0$ is fixed at an arbitrary $x \in S$. The dynamics function $f : S \times A \times D \rightarrow S$ is Borel measurable. Given $(X_t,U_t)$, the disturbance $W_t$ is conditionally independent of $W_s$ for all $s \neq t$, and the distribution of $W_t$ is known.\footnote{Precise knowledge of the disturbance distributions and the dynamics function $f$ are limitations of this standard formulation. If these assumptions are too strong for one's application of interest, then it may be appropriate to consider a \emph{distributionally robust} formulation, for instance, see \cite{van2015distributionally}, which studies a risk-sensitive linear-quadratic setting. Online estimation of $f$ is a growing research area, e.g., see \cite{onlinelearning1, onlinelearning2, onlinelearning3} for some recent works.} If $(x_t,u_t) \in S \times A$ is the realization of $(X_t,U_t)$, then the distribution of $W_t$ is $p(\cdot|x_t,u_t)$.\footnote{$p(\cdot|\cdot,\cdot)$ is a \emph{Borel-measurable stochastic kernel} on $D$ given $S \times A$ \cite[Def. 7.12, p. 134]{bertsekas2004stochastic}, whose meaning we explain next. Let $\mathcal{P}(D)$ denote the space of probability measures on $(D,\mathcal{B}_{D})$ with the weak topology \cite[p. 127]{bertsekas2004stochastic}. The function $\psi : S \times A \rightarrow \mathcal{P}(D)$ such that $\psi(x,u) := p(\cdot|x,u)$ is Borel measurable, i.e., measurable relative to $\mathcal{B}_{S \times A}$ and $\mathcal{B}_{\mathcal{P}(D)}$. Later, to guarantee the existence of an optimal policy, we assume that $\psi$ is continuous, which holds if $p(\cdot|x,u)$ is constant in $(x,u)$, for example.\label{footnote9}} 

The random cost $Z : \Omega \rightarrow \mathbb{R}$ is a Borel-measurable function, whose domain is a sample space $\Omega := (S \times A)^N \times S$. Any $\omega = (x_0,u_0,\dots,x_{N-1},u_{N-1},x_N) \in \Omega$ is a realization of the random trajectory $(X_0,U_0,\dots,X_{N-1},U_{N-1},X_N)$. In particular, $Z$ takes the form, for any $\omega = (x_0,u_0,\dots,x_{N-1},u_{N-1},x_N) \in \Omega $,
\begin{equation}\label{myZ}
  \textstyle  Z(\omega) := c_N(x_N) + \sum_{t = 0}^{N-1} c(x_t,u_t) \geq \underline{b}.
\end{equation}
The stage cost $c : S \times A \rightarrow \mathbb{R}$ and the terminal cost $c_N : S \rightarrow \mathbb{R}$ are Borel measurable and bounded below by $\underline{d} \in \mathbb{R}$, and $Z$ is bounded below by $\underline{b} := (N+1)\underline{d}$. For convenience, we define a non-negative random cost $Z'$,
\begin{equation}\label{zprime}\begin{aligned}
   \textstyle  Z'(\omega) & := c_N'(x_N) + \textstyle \sum_{t=0}^{N-1} c'(x_t,u_t)  = Z(\omega) - \underline{b},
\end{aligned}\end{equation}
where $c' := c - \underline{d}$ and $c'_N := c_N - \underline{d}$ are translated versions of $c$ and $c_N$, respectively. 

We consider two classes of control policies: $\Pi$ is a class of history-dependent policies, and $\Pi'$ is the class of deterministic Markov policies. In particular, any $\pi \in \Pi'$ takes the form $\pi = (\mu_0,\mu_1,\dots,\mu_{N-1})$, where $\mu_t : S \rightarrow A$ is Borel measurable for all $t \in \mathbb{T}$. We present $\Pi$ in detail in Sec. \ref{exactmethodcvar}.

The risk-neutral approach for managing the uncertainty in $Z$ is to minimize the expectation $E_x^\pi(Z)$ over the class of policies $\Pi'$. If $G : \Omega \rightarrow \mathbb{R}$ is Borel measurable, then $E_x^\pi(G) := \int_\Omega G \; \mathrm{d}P_x^\pi$ is the expectation of $G$ with respect to $P_x^\pi$, a probability measure on $(\Omega,\mathcal{B}_\Omega)$. $P_x^\pi$ depends on an initial condition $x$ and a policy $\pi$ and provides the probabilities of the states and controls being in Borel-measurable subsets of $S$ and $A$, respectively \cite[pp. 190--191]{bertsekas2004stochastic} \cite[p. 16]{hernandez2012discrete}. 

In contrast to the risk-neutral approach, we study two risk-averse approaches, where we aim to minimize the EU of $Z$ or the CVaR of $Z$. We require some conditions to ensure that the optimal values of the problems of interest are finite.
\begin{assumption}\label{assumption11}
We assume the following conditions:
\begin{enumerate}
    \item Let $\Theta \subseteq (-\infty,0)$ be non-empty. For all $\theta \in \Theta$, there is a $\pi_\theta \in \Pi'$ s.t. $E_x^{\pi_\theta }(e^{\frac{-\theta}{2}Z'}) < +\infty$ for all $x \in S$.
    \item There is a $\pi \in \Pi$ s.t. $E_x^\pi(|Z|) < +\infty$ for all $x \in S$.
\end{enumerate}
\end{assumption}
For example, if $c$ and $c_N$ are bounded, then Assumption \ref{assumption11} holds.
Next, we introduce the EU and CVaR functionals.
\subsection{Exponential Utility (EU)}\label{euintro}
The EU functional assesses larger values of a random cost through an exponential transformation that depends on a parameter $\theta \in \Theta$.
For any  $x \in S$ and $\pi \in \Pi'$, the EU of $Z$ at level $\theta \in \Theta$ is given by
\begin{align}\label{exputility}
    \rho_{\theta,x}^\pi(Z) \hspace{-.5mm} := \hspace{-.2mm} \underline{b} + \textstyle\frac{-2}{\theta}\log E_x^\pi\big(e^{\frac{-\theta}{2} Z'}\big)\hspace{-.5mm} = \hspace{-.5mm} \textstyle\frac{-2}{\theta}\log E_x^\pi\big(e^{\frac{-\theta}{2} Z}\big).
\end{align}
We define $\rho_{\theta,x}^\pi(Z)$ so that $\frac{-\theta}{2}>0$ multiplies $Z'$, which is a non-negative random variable. If $\theta$ is more negative, then 
larger values of $Z'$ are considered to be more harmful, and therefore, more critical to assess when synthesizing a policy. Hence, a more negative value of $\theta$ represents a higher degree of risk aversion. %
The equality in \eqref{exputility} holds as a consequence of $Z = \underline{b}+Z'$.
It can be shown under certain conditions that $\underset{\theta \rightarrow 0}{\lim} \rho_{\theta,x}^\pi(Z) = \underline{b} + E_x^\pi(Z')$. If $|\theta|$ is sufficiently small and if $\{ E_x^\pi((Z')^n) : n \in \mathbb{N}\}$ is bounded, then the EU of $Z$ approximates a weighted sum of the expectation $E_x^\pi(Z)$ and variance $\text{var}_x^\pi(Z)$,
\begin{equation}\begin{aligned}\label{approx2}
    \rho_{\theta,x}^\pi(Z)  \hspace{-.2mm} \approx \hspace{-.2mm} \underline{b} + E_x^\pi(Z') - \textstyle\frac{\theta}{4}\text{var}_x^\pi(Z') 
     \hspace{-.2mm} = \hspace{-.2mm} E_x^\pi(Z) - \textstyle\frac{\theta}{4}\text{var}_x^\pi(Z).
\end{aligned}\end{equation}
For details regarding the limit result or \eqref{approx2}, please refer to \cite[p. 765]{whittle1981} or the supplementary material.

\subsection{Conditional Value-at-Risk (CVaR)}\label{cvarintro}
CVaR uses quantiles rather than a transformation to assess larger realizations of a random cost. Let $x \in S$ and $\pi \in \Pi$ be given. We denote the CVaR of $Z$ with respect to $P_x^\pi$ at level $\alpha \in (0,1]$ by $\text{CVaR}_{\alpha,x}^\pi(Z)$.\footnote{For the definition of $P_x^\pi$ in this context, please refer to the supplementary material (p. 4).} $\text{CVaR}_{\alpha,x}^\pi(Z)$ can be written in terms of the left-side $(1-\alpha)$-quantile of the distribution of $Z$. This quantile is called the \emph{Value-at-Risk} of $Z$ at level $\alpha \in (0,1)$, which is defined by
\begin{equation}
    \text{VaR}_{\alpha,x}^\pi(Z) := \inf\big\{z \in \mathbb{R} : P_x^\pi\big(\{Z\leq z\}\big) \geq 1-\alpha\big\},
\end{equation}
where $F_{Z,x}^\pi(z) := P_x^\pi\big(\{Z\leq z\}\big)$ is the CDF of $Z$ for the initial condition $x$ and policy $\pi$.\footnote{Another name for $\text{VaR}_{\alpha,x}^\pi(Z)$ is the generalized inverse CDF of $Z$ at level $1-\alpha$.} 
$\text{CVaR}_{\alpha,x}^\pi(Z)$ is the expectation of $Z$ conditioned on the event $\{Z \geq \text{VaR}_{\alpha,x}^\pi(Z)\}$, if $\alpha \in (0,1)$, $F_{Z,x}^\pi$ is continuous at $z = \text{VaR}_{\alpha,x}^\pi(Z)$, and $E_x^\pi(|Z|)$ is finite \cite[Thm. 6.2]{shapiro2009lectures}. This fact motivates the name \emph{Conditional Value-at-Risk}. In this setting, $\text{CVaR}_{\alpha,x}^\pi(Z)$ is the expectation of $Z$ in the $\alpha \cdot 100\%$ worst cases. The parameter $\alpha$ is a risk-aversion level that represents a fraction of the largest values of $Z$ that are of particular concern. Another name for CVaR is \emph{Average Value-at-Risk} because
\begin{equation}
    \text{CVaR}_{\alpha,x}^\pi(Z) = \textstyle \frac{1}{\alpha} \int_{1-\alpha}^1 \text{VaR}_{1-\ell,x}^\pi(Z) \; \mathrm{d}\ell,
\end{equation}
provided that $E_x^\pi(|Z|) < +\infty$ and $\alpha \in (0,1]$; see \cite[Thm. 6.2]{shapiro2009lectures} for a proof in the case of $\alpha \in (0,1)$.

As outlined above, various representations for CVaR are used in the literature. The next representation, which often serves as the definition for CVaR \cite{shapiro2012}, is convenient for optimal control problems in particular, e.g., see \cite{milleryang, bauerleott, chapmantac2022}. For any $\alpha \in (0,1]$, if $E_x^\pi(|Z|) < +\infty$, then the $\text{CVaR}_{\alpha,x}^\pi(Z)$ is defined by
\begin{equation}\label{cvardef}
    \text{CVaR}_{\alpha,x}^\pi(Z) := \inf_{s \in \mathbb{R}} \big( s + \textstyle\frac{1}{\alpha} E_x^\pi(\max\{Z-s,0\}) \big).
\end{equation}
If $E_x^\pi(|Z|) = +\infty$, then $\text{CVaR}_{\alpha,x}^\pi(Z) := +\infty$. Note that $\text{CVaR}_{1,x}^\pi(Z)$ equals $E_x^\pi(Z)$. Eq. \eqref{cvardef} allows us to write $\text{CVaR}_{\alpha,x}^\pi(Z)$ as a weighted sum of $\text{VaR}_{\alpha,x}^\pi(Z)$ and the expectation of how much $Z$ exceeds $\text{VaR}_{\alpha,x}^\pi(Z)$. Precisely, if $E_x^\pi(|Z|) < +\infty$ and $\alpha \in (0,1)$, then a minimizer of the objective function in \eqref{cvardef} is $\text{VaR}_{\alpha,x}^\pi(Z)$ \cite{shapiro2012}, and thus,
\begin{equation}\label{howencoderiskcvar}
  \text{CVaR}_{\alpha,x}^\pi(Z) = \text{VaR}_{\alpha,x}^\pi(Z) + \textstyle\frac{1}{\alpha} E_x^\pi(\max\{Z-\text{VaR}_{\alpha,x}^\pi(Z),0\}).
\end{equation}

\subsection{Comparison between Exponential Utility and CVaR}\label{comp}
Table \ref{tablecomp} summarizes how EU and CVaR encode risk aversion differently. 
The approximation \eqref{approx2} suggests that EU encodes risk aversion in terms of the spread of the distribution of $Z$ relative to the moment $E_x^\pi(Z)$. Eq. \eqref{howencoderiskcvar} indicates that CVaR encodes risk aversion in terms of the expected exceedance of $Z$ relative to the quantile $\text{VaR}_{\alpha,x}^\pi(Z)$.

\begin{table}[h]
\caption{How EU and CVaR Encode Risk Aversion}
\label{table}
\setlength{\tabcolsep}{3pt}
\begin{tabular}{|p{60pt}|p{85pt}|p{90pt}|}
\hline
 &  \textbf{Exponential Utility (EU) \eqref{exputility}} & \textbf{Conditional Value-at-Risk (CVaR) \eqref{cvardef}}  \\
 \hline
\textbf{Approach} & Quantifies a distribution in terms of its moments. & Quantifies a distribution in terms of its quantiles.\\
\hline
\textbf{Interpretation} & Approximates a weighted sum of mean and variance if $|\theta|$ is small. & Approximates an expectation in a fraction of worst cases. \\
\hline
\textbf{Parameter} & $\theta \in \Theta \subseteq (-\infty,0)$ & $\alpha \in (0,1]$\\
\hline
\textbf{More risk-averse} & $\theta$ is more negative. & $\alpha$ is near 0.\\
\hline
\textbf{Less risk-averse} & $\theta$ is near 0. &  $\alpha$ is near 1. \\
\hline 
\end{tabular}
\label{tablecomp}
\end{table}

In the next section, we present optimal control problems, in which we define the objective functions in terms of EU and CVaR. Subsequently, we solve these problems numerically for a thermostatic regulator and for a stormwater system. Then, we simulate trajectories under a policy that has been optimized numerically with respect to EU or CVaR. We use such simulations to estimate an optimal distribution of $Z$ and to examine trade-offs that arise in the empirical statistics of $Z$, as the risk aversion level varies.

\section{Algorithms for Optimal Control with\\ Exponential Utility and CVaR Objectives}\label{SecIII}
We present two distinct approaches for risk-averse optimal control.
\begin{problem}[EU-Optimal Control]\label{prob1}
Consider an optimal control problem in which the EU at level $\theta \in \Theta$ is used to assess $Z$,
    $V_\theta^*(x) := \inf_{\pi \in \Pi'} \rho_{\theta,x}^\pi(Z)$ for all $x \in S$.
If there is a $\pi_\theta^* \in \Pi'$ such that $V_\theta^*(x) = \rho_{\theta,x}^{\pi_\theta^*}(Z)$ for all $x \in S$, $\pi_\theta^*$ is said to be \emph{optimal for $V_\theta^*$}. 
\end{problem}

\begin{problem}[CVaR-Optimal Control]\label{cvarproblem}
Consider an optimal control problem in which the CVaR at level $\alpha \in (0,1]$ is used to assess $Z$, $J_\alpha^*(x) := \inf_{\pi \in \Pi} \text{CVaR}_{\alpha,x}^\pi(Z)$ for all $x \in S$. $\Pi$ is a class of history-dependent policies, which we specify formally later in this section. If there is a $\pi_\alpha^* \in \Pi$ such that $J_\alpha^*(x) = \text{CVaR}_{\alpha,x}^{\pi_\alpha^*}(Z)$ for all $x \in S$, $\pi_\alpha^*$ is said to be \emph{optimal for $J_\alpha^*$}. 
\end{problem}
\begin{remark}[Finiteness of optimal value functions]
The first (second) condition of Assumption \ref{assumption11} and $Z$ being bounded below by $\underline{b} \in \mathbb{R}$ imply that $V_\theta^*$ ($J_\alpha^*$) is finite.
\end{remark}

A dynamic programming (DP) algorithm on the state space $S$ can be defined to compute $V_\theta^*$ under some conditions (to be presented). An algorithm is provided by \cite[Prob. 7(b), p. 66]{Bertsekas1976} in a setting without $\theta$, for example. However, CVaR does not satisfy a DP recursion on $S$, and thus, computing $J_\alpha^*$ requires more complicated algorithms. Next, we present a method to compute $V_\theta^*$ exactly in principle. Then, we present different approaches to estimate $J_\alpha^*$ or compute $J_\alpha^*$ exactly in principle.

\subsection{Dynamic Programming for EU-Optimal Control}
The following assumption permits the exact computation of $V_\theta^*$ in principle and guarantees the existence of a policy that is optimal for $V_\theta^*$.
\begin{assumption}[Measurable Selection Condition]\label{measselect} \hphantom{Assume}
\begin{enumerate}
\item The distribution of $W_t$, $p(\cdot|\cdot,\cdot)$, is a continuous stochastic kernel on $D$ given $S \times A$ (Footnote \ref{footnote9}).
    \item The dynamics function $f$ is continuous. $c$ and $c_N$ are l.s.c. and bounded; i.e., $\underline{d}\leq c \leq \bar{d}$ and $\underline{d}\leq c_N \leq \bar{d}$.
    \item The set of controls $A$ is compact.
\end{enumerate}
\end{assumption}
\begin{remark}[Justification of Assumption \ref{measselect}]
Assumption \ref{measselect} is a measurable selection condition. Such conditions are used for stochastic control problems on continuous spaces, in which the costs are non-quadratic or the dynamics function is non-linear. The conditions guarantee the existence of an optimal policy. For additional examples, please see \cite[Sec. 2, p. 106]{bauerlerieder}, \cite[Def. 8.7, pp. 208--209]{bertsekas2004stochastic}, and \cite[Sec. 3.3, pp. 27--29]{hernandez2012discrete}. 

\end{remark}

Next, we provide a DP algorithm for $V_\theta^*$.
\begin{algorithm}[Exact DP for $V_\theta^*$]\label{valalgwhittle}
For any $\theta \in \Theta$, define the functions $V_{N}^\theta,\dots,V_{1}^\theta,V_0^\theta$ on $S$ recursively as follows: for all $x \in S$, $V_{N}^\theta(x) := c_N(x)$, and for $t = N-1, \dots,1, 0$, $V_{t}^\theta(x) := \inf_{u \in A} v_{t+1}^\theta(x,u)$, where
    $ v_{t+1}^\theta(x,u) 
     :=  c(x,u) + 
 {\textstyle\frac{-2}{\theta}} \log \textstyle (\int_{D} \exp(\frac{-\theta}{2} V_{t+1}^\theta(f(x,u,w))) p(\mathrm{d}w|x,u))$
for all $(x,u) \in S \times A$.
\end{algorithm}

We have studied Alg. \ref{valalgwhittle} formally in \cite{classicalrstheorypaper}, and we summarize the analysis next. Under Assumption \ref{measselect}, for all $t \in \mathbb{T}'$, $V_{t}^\theta$ is l.s.c. and bounded, and for all $t \in  \mathbb{T}$, there is a Borel-measurable function $\mu_{t}^\theta : S \rightarrow A$ such that $V_{t}^\theta(x) = v_{t+1}^\theta(x,\mu_{t}^\theta(x))$ for all $x \in S$. The policy $\pi_\theta^*:= (\mu_{0}^\theta, \mu_{1}^\theta, \dots, \mu_{N-1}^\theta)$ satisfies $V_\theta^*(x) = \rho_{\theta,x}^{\pi_\theta^*}(Z) = V_0^\theta(x)$ for all $x \in S$, and in particular, $\pi_\theta^*$ is optimal for $V_\theta^*$.
\begin{remark}[Alg. \ref{valalgwhittle} restricted to non-negative costs]\label{remarkmod}
It is common to define the DP iterates $V_{N}'^\theta,\dots,V_{1}'^\theta,V_0'^\theta$ in terms of the non-negative costs $c_N'$ and $c'$. Similarly in this setting, under Assumption \ref{measselect}, there is a policy $\pi_\theta^* \in \Pi'$ such that $V_\theta^*(x) = \rho_{\theta,x}^{\pi_\theta^*}(Z) = \underline{b} + V_0'^\theta(x)$ for all $x \in S$.
\end{remark}

In this paper, we numerically investigate Alg. \ref{valalgwhittle} with and without the restriction to non-negative costs. In Sec. \ref{numresults_optimal_control}, we show that the unrestricted form of Alg. \ref{valalgwhittle} permits a larger range of numerically stable values of $\theta$.

In the absence of quadratic costs, linear dynamics, and an analytical expression for the disturbance kernel $p(\cdot|\cdot,\cdot)$, one typically implements a DP algorithm numerically. While the details of such implementations are problem-dependent, a standard implementation involves discretizing continuous spaces and interpolating values that are computed on discrete spaces. That is, the state space $S$ and the control space $A$ may be replaced by non-empty subsets $S_G \subseteq S$ and $A_G \subseteq A$, respectively, containing finitely many elements. A discrete approximate distribution for the disturbance may be estimated from observations. To improve the efficiency of DP algorithms, approximate methods are being studied, including stochastic rollout, e.g., see \cite{Esfahani, bertrein}, and the references therein for details regarding the state-of-the-art. A related line of research has demonstrated the efficacy of using approximate, efficient ``warm-start'' computations to estimate high-fidelity, computationally expensive computations; e.g., see \cite{kenethesis} for a robust setting with continuous-time non-stochastic systems and our work \cite[Sec. VI]{chapmantac2022} for a risk-averse setting with discrete-time stochastic systems.
Next, we present different approaches for solving or approximating Problem \ref{cvarproblem}.%
\subsection{Approaches for CVaR-Optimal Control}
\subsubsection{Brute Force Simulation}
A brute force approach to estimate $J_\alpha^*(x) := \inf_{\pi \in \Pi} \text{CVaR}_{\alpha,x}^\pi(Z)$ is to estimate a distribution for $Z$ for a given policy $\pi$ and initial condition $x$ via Monte Carlo simulations. Then, one can use a CVaR estimator, e.g., \cite[p. 300]{shapiro2009lectures}, to estimate $\text{CVaR}_{\alpha,x}^\pi(Z)$. One can repeat this procedure for many policies and any initial conditions of interest. However, it is generally not clear which policies one should simulate, and simulating all policies is seldom feasible. Fortunately, alternate approaches are available.
\subsubsection{Exact Method}\label{exactmethodcvar}
B\"{a}uerle and Ott developed a method to compute $J_\alpha^*$ exactly in principle \cite{bauerleott}. The key machinery in \cite{bauerleott} is to define the dynamics of an extra state that records a cumulative cost from time zero to the current time. A DP algorithm is defined on the augmented state space, and a deterministic Markov policy on the augmented state space is shown to be optimal under a measurable selection condition \cite{bauerleott}. In the setting of a finite time horizon, the range of the extra state depends on the largest value of the cumulative cost, which limits computational tractability. Our numerical examples (to be shown in Sec. \ref{numresults_optimal_control}) demonstrate this reduction in tractability compared to EU-optimal control, which does not require an extra state. Next, we present the method of \cite{bauerleott} to compute $J_\alpha^*$ for the control system model of Sec. \ref{controlsysmodel}.
\begin{enumerate}
    \item \emph{Define the dynamics of the extra state $S_t$.} The augmented state is $(X_t,S_t)$. The augmented state space is $S \times \mathcal{Z}$, where $\mathcal{Z} := [-\bar{a},\bar{a}] \subseteq \mathbb{R}$ and $0 \leq Z' \leq \bar{a} := (\bar{d}-\underline{d})(N+1)$ everywhere. For any $t \in \mathbb{T}$, if $(x_t,s_t,u_t) \in S \times \mathcal{Z} \times A$ is the realization of $(X_t,S_t,U_t)$, then the realization of $S_{t+1}$ is $s_{t+1} = s_t - c'(x_t,u_t)$.
    \item \emph{Define a class of policies $\Pi$ that are history-dependent through $(X_t,S_t)$.} Any $\pi \in \Pi$ takes the form $\pi = (\pi_0,\pi_1,\dots,\pi_{N-1})$, where $\pi_t(\cdot|\cdot,\cdot)$ is a Borel-measurable stochastic kernel on $A$ given $S \times \mathcal{Z}$.\footnote{Given a policy $\pi \in \Pi$, if $(x_t,s_t) \in S \times \mathcal{Z}$ is the realization of $(X_t,S_t)$, then $\pi_t(\cdot|x_t,s_t) \in \mathcal{P}(A)$ is the distribution of $U_t$, where $\mathcal{P}(A)$ is the collection of probability measures on $(A,\mathcal{B}_A)$. The map $\psi : S \times \mathcal{Z} \rightarrow \mathcal{P}(A)$ such that $\psi(x_t,s_t) := \pi_t(\cdot|x_t,s_t)$ is Borel measurable.}
    \item \emph{Define a CVaR-optimal control problem for $Z'$.} Define $W_\alpha^*(x):= \inf_{\pi \in \Pi} \text{CVaR}_{\alpha,x}^\pi(Z')$ for all $x \in S$. Since $Z' = Z - \underline{b}$ and $Z$ is bounded everywhere, it holds that $W_\alpha^* = J_\alpha^* - \underline{b}$ for any $\alpha \in (0,1]$.\footnote{We use translation-equivariance of $\text{CVaR}$. That is, if $Y$ is a random variable with finite expectation, $a \in \mathbb{R}$, and $\alpha \in (0,1]$, then $\text{CVaR}_\alpha(Y + a) = \text{CVaR}_\alpha(Y) + a$.\label{transeqvar}} \label{mystep3} 
    \item \emph{Re-express $W_\alpha^*$ into a useful form for computation.} By the definition of $\text{CVaR}$ \eqref{cvardef} and $0\leq Z'\leq \bar{a}$ everywhere, it holds that, for all $x \in S$ and $\alpha \in (0,1]$,
    \begin{align*}
        W_\alpha^*(x) & = \inf_{s \in \mathbb{R}} \big( s + {\textstyle \frac{1}{\alpha}} \inf_{\pi \in \Pi} E_x^\pi(\max\{Z' - s,0 \}) \big) \nonumber\\
        & = \min_{s \in [0, \bar{a}]} \big( s + {\textstyle \frac{1}{\alpha}} \inf_{\pi \in \Pi} E_x^\pi(\max\{Z' - s,0 \}) \big),
    \end{align*}
    where a minimizer $s_{\alpha,x}^* \in [0, \bar{a}]$ exists for the outer optimization problem \cite[Lemma 1]{chapmantac2022}.
    We define the inner optimization problem for any $(x,s) \in S \times \mathbb{R}$ by
    \begin{equation}\label{innerproblem}
        J^*(x,s) := \inf_{\pi \in \Pi} E_x^\pi(\max\{Z' - s,0 \})
    \end{equation}
    and simplify $W_\alpha^*$ as follows:
    \begin{equation}\label{WandV0}
        W_\alpha^*(x) \hspace{-.6mm} = \hspace{-.8mm} \min_{s \in [0, \bar{a}]} \big( s + {\textstyle \frac{1}{\alpha}} J^*(x,s) \big)\hspace{-.6mm} =\hspace{-.6mm} s_{\alpha,x}^* + {\textstyle \frac{1}{\alpha}} J^*(x,s_{\alpha,x}^*).
    \end{equation}
    \item \emph{Use DP on $S \times \mathcal{Z}$ to compute $J^*$.} Algorithm \ref{dpcvar}, to be presented, defines a function $J_t$ on $S \times \mathcal{Z}$ recursively for $t= N,\dots,1,0$, and under appropriate conditions, it holds that $J_0 = J^*$ (Thm. \ref{thmdpcvar}, to be presented).
 \item \emph{Use $J^*$ to compute $J_\alpha^*$.} For any $\alpha \in (0,1]$ and $x \in S$, compute $W_\alpha^*(x)$ using \eqref{WandV0}. 
 Then, using step \ref{mystep3}, compute $J_\alpha^* = W_\alpha^* + \underline{b}$. %
\end{enumerate}

Next, we present Algorithm \ref{dpcvar} and Theorem \ref{thmdpcvar}.
\begin{algorithm}[Exact DP for $J^*$]\label{dpcvar}
Define the functions $J_N, \dots, J_1, J_0$ on $S \times \mathcal{Z}$ recursively as follows: for any $(x,s) \in S \times \mathcal{Z}$, $J_N(x,s) := \max\{c_N'(x) -s ,0 \}$, and for $t = N-1,\dots,1,0$, $J_t(x,s)  := \inf_{u \in A} v_{t+1}(x,s,u)$, where $v_{t+1}$ is defined by\vspace{-1mm}
\begin{equation*}\label{vtplus1}
    v_{t+1}(x,s,u) := \textstyle \int_D J_{t+1}(f(x,u,w), s - c'(x,u)) \; p(\mathrm{d}w|x,u)\vspace{-1mm}
\end{equation*}
with $(x,s,u) \in S \times \mathcal{Z} \times A$.
\end{algorithm}

The following theorem provides an analysis of Algorithm \ref{dpcvar}. In the interest of space, technical details and the proof are omitted. Similar proof techniques can be found in \cite{bertsekas2004stochastic}, \cite{hernandez2012discrete}, \cite{bauerleott}, and \cite{chapmantac2022}.
\begin{theorem}[Analysis of Alg. \ref{dpcvar}]\label{thmdpcvar}
Let Assumption \ref{measselect} hold, and assume that $c$ and $c_N$ are continuous. Then, $J_t$ is bounded and continuous for each $t \in \mathbb{T}'$. For each $t \in \mathbb{T}$, there is a Borel-measurable function $\kappa_t : S \times \mathcal{Z} \rightarrow A$ such that $J_t(x,s) = v_{t+1}(x,s,\kappa_t(x,s))$ for all $(x,s) \in S \times \mathcal{Z}$. Define $\pi^* := (\kappa_0,\kappa_1,\dots,\kappa_{N-1})$. It holds that $J_0(x,s) = J^*(x,s) = E_x^{\pi^*}(\max\{Z' - s,0 \})$ for all $(x,s) \in S \times \mathcal{Z}$.
\end{theorem}

\begin{remark}[Policy deployment on $S \times \mathcal{Z}$]
Thm. \ref{thmdpcvar} guarantees the existence of a policy $\pi^* \in \Pi$ such that for all $x \in S$ and $\alpha \in (0,1]$,
\begin{align}
    J_\alpha^*(x) %
     &= \underline{b} + \inf_{s \in \mathbb{R}} \big( s + {\textstyle \frac{1}{\alpha}} E_x^{\pi^*}(\max\{Z' - s,0 \})\big) \\
     &= \underline{b} + \text{CVaR}_{\alpha,x}^{\pi^*}(Z')\\
     & = \text{CVaR}_{\alpha,x}^{\pi^*}(Z),
\end{align}
and thus, $\pi^*$ is optimal for $J_\alpha^*$. Next, we explain how to deploy $\pi^* = (\kappa_0,\kappa_1,\dots,\kappa_{N-1})$. %
Let $x \in S$ and $\alpha \in (0,1]$ be given, and let $s_{\alpha,x}^* \in [0, \bar{a}]$ satisfy \eqref{WandV0}. The realization of $(X_0,S_0)$ is $(x_0,s_0)=(x, s_{\alpha,x}^*)$. For $t = 0,1,\dots,N-1$, repeat the following steps:
\begin{enumerate}
    \item Choose the control $u_t = \kappa_t(x_t,s_t)$.
    \item A realization $w_t$ of $W_t$ occurs according to the distribution $p(\cdot|x_t,u_t)$.
    \item Then, the realization of $(X_{t+1},S_{t+1})$ is given by $(x_{t+1},s_{t+1}) = (f(x_t,u_t,w_t),s_t - c'(x_t,u_t))$. 
    \item Update $t$ by 1, and proceed to step 1 if $t < N$.
\end{enumerate}
\end{remark}

Subsequently, we describe alternative approaches for estimating $J_\alpha^*$.
\subsubsection{Upper-Bound Method}\label{upperboundmethod}
We derive an upper bound for $J_\alpha^*$ by adopting techniques from our prior work \cite{chapmantac2021}.\footnote{In this prior work, we derived an upper bound for the CVaR of a maximum cost to approximate a safety analysis problem \cite{chapmantac2021}. However, in the current work, we consider a cumulative cost $Z$.} The upper bound can be computed by DP on $S$ and therefore has the benefit of providing a theoretically-guaranteed estimate of $J_\alpha^*$ without state-space augmentation. 

Assume that $c$ and $c_N$ are l.s.c. and bounded below by $\underline{d}$. Recall that $Z' = Z - \underline{b} \geq 0$ is given by \eqref{zprime}, and $\Pi'$ is the class of deterministic Markov policies on $S$.
Let $x \in S$, $\alpha \in (0,1]$, and $\pi \in \Pi'$ be given. First, suppose that $E_x^\pi(Z')$ is finite. Then, $ \text{CVaR}_{\alpha,x}^\pi(Z') \leq \frac{1}{\alpha}E_x^\pi(Z')$ by \cite[Lemma 2]{chapmantac2021} and $\text{CVaR}_{\alpha,x}^\pi(Z') = \text{CVaR}_{\alpha,x}^\pi(Z) - \underline{b}$ by translation-equivariance (Footnote \ref{transeqvar}).
It follows that $\text{CVaR}_{\alpha,x}^\pi(Z) \leq \underline{b} + \frac{1}{\alpha}E_x^\pi(Z')$.
Otherwise, if $E_x^\pi(Z') = +\infty$, then $E_x^\pi(|Z|) = +\infty$, and 
then $\text{CVaR}_{\alpha,x}^\pi(Z) = +\infty$ by definition (Sec. \ref{cvarintro}). Thus, regardless of whether $E_x^\pi(Z')$ is finite, it holds that
 $\text{CVaR}_{\alpha,x}^\pi(Z) \leq \underline{b} + \frac{1}{\alpha}E_x^\pi(Z')$ $\forall x \in S \; \forall \pi \in \Pi' \; \forall \alpha \in (0,1]$, and hence,
\begin{equation}\label{my23}
 J_\alpha^*(x) \leq \inf_{\pi \in \Pi'} \text{CVaR}_{\alpha,x}^\pi(Z) \leq \underline{b} + {\textstyle \frac{1}{\alpha}}\inf_{\pi \in \Pi'} E_x^\pi(Z').
\end{equation}
for all $x \in S$ and $\alpha \in (0,1]$.
One may compute $J'(x) := \inf_{\pi \in \Pi'} E_x^\pi(Z')$ for all $x \in S$ exactly in principle under appropriate conditions, e.g., see \cite{hernandez2012discrete, bertsekas2004stochastic}, and then use $J'$ to upper-bound $J_\alpha^*$ for any $\alpha \in (0,1]$ of interest \eqref{my23}. 
\subsubsection{Approximate Method}\label{chowmethod}
Chow et al. proposed a method to approximate $J_\alpha^*$ using linear programming and an extra state that takes values in $(0,1]$ \cite{chow2015risk}. The extra state may be interpreted as a time-varying risk-aversion level. The method in \cite{chow2015risk} assumes that the CVaR of a random cumulative cost initialized at time $t$ is well-approximated by the optimal value of a linear program (LP). The LP is inspired by a history-dependent temporal decomposition for CVaR \cite[Thm. 6]{pflug2016timeEuro} \cite[Thm. 21, Lemma 22]{pflug2016timeMath}. The decomposition is related to a representation for CVaR, which takes the form of a distributionally robust expectation \cite[Eqn. 3.5, Eqn. 3.13]{shapiro2012}. The method in \cite{chow2015risk} defines the iterates of a DP recursion to resemble the decomposition but to be history-dependent only through the current augmented state. While an upper or lower bound to a CVaR-optimal control problem is not guaranteed by \cite{chow2015risk}, the approach is intriguing, as the additional state has a small range. However, numerically solving the required LPs may be difficult in practice because the constraints may span several orders of magnitude. Solving many small LPs in parallel can be used to overcome the difficulty of differently scaled constraints that can arise in large LPs.

\section{System Models}\label{sysmodels}
We consider two control systems, a thermostatic regulator and a stormwater system. For each system, we choose a cost function $c = c_N$ that depends on the current state so that $Z$ represents a cumulative deviation of the state trajectory relative to a particular subset $K$ of  $S$.\footnote{We formalize the meaning of deviation later in this section. One may choose $c$ to depend on the current state and control, in which case, $Z$ would have a different interpretation.} The desired outcome is that the state trajectory remains inside of $K$ over time. However, this may not be possible due to disturbances that arise in the environment. Thus, we permit departures from $K$, and we choose $c$ to quantify the magnitude of departure. %
That is, $K$ is a soft safety constraint set that characterizes a desired operating region.
\subsection{Thermostatic Regulator}
\begin{figure}[t]
\centerline{\includegraphics[trim={0.45cm 0 0.75cm 0},clip,width=\columnwidth]{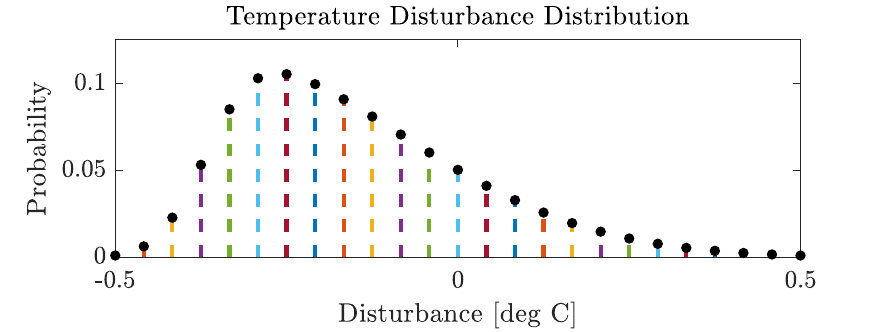}}
\caption{A right-skewed distribution for the disturbances in the thermostatic regulator model.}
\label{temperature_disturbance}
\end{figure}

The temperature of a room equipped with a heater that is regulated by a thermostat may be modeled as follows \cite{yang2018dynamic, mort}: for $t = 0,1,\dots,N-1$,
   $ x_{t+1} = a x_t + (1-a) (b - \eta R P u_t) + w_t$.
$x_t \in \mathbb{R}$ ($^{\circ}\text{C}$) is a value of the random temperature $X_t$. $u_t \in [0,1]$ is a value of $U_t$, representing an amount of power supplied to the system. $w_t \in \mathbb{R}$ is a value of $W_t$, which represents environmental uncertainties. Table \ref{paramtemp} lists the model parameters. We consider a temperature disturbance distribution with positive (i.e, right) skew (Fig. \ref{temperature_disturbance}).

We define $c$ to quantify a deviation relative to a desired temperature range $K = [20, 21]$ $^{\circ}\text{C}$ as follows:
   $ c(x,u) := c_N(x) := \max\{ x - 21, 20 - x\}$
for any $(x,u) \in \mathbb{R} \times [0,1]$. The random cumulative cost $Z$ \eqref{myZ} is the total deviation of the random state trajectory relative to the desired temperature range.

\begin{table}
\caption{Thermostatic Regulator Parameters}
\label{table}
\setlength{\tabcolsep}{3pt}
\begin{tabular}{|p{25pt}|p{100pt}|p{77pt}|}
\hline
\textbf{Symbol} &  \textbf{Description} & \textbf{Value}  \\
\hline
$a$ & Time delay & $e^{\frac{-\triangle \tau}{CR}}$ (no units) \\
$b$ & Temperature shift & 32 $^{\circ}\text{C}$ \\
$C$ & Thermal capacitance & 2 $\frac{\text{kWh}}{^{\circ}\text{C}}$ \\
$\eta$ & Control efficiency & 0.7 (no units) \\
$K$ & Constraint set & $[20, 21]$ $^{\circ}\text{C}$ \\
$P$ & Power & 14 kW \\
$R$ & Thermal resistance & 2 $\frac{^{\circ}\text{C}}{\text{kW}}$ \\
$\triangle \tau$ & Duration of $[t, t+1)$ & $\frac{5}{60}$ h \\
$N$ & Number of time points & 12 (= 1 h) \\
$A_G$ & Grid of controls & $\{0,0.1,\dots,1\}$ (no units)\\
$S_G$ & Grid of states & $\{18, 18.1, \dots, 23\}$ $^{\circ}\text{C}$ \\
\hline
\multicolumn{3}{p{242pt}}{h $=$ hours, kW $=$ kilowatts, $^{\circ}\text{C}$ $=$ degrees Celsius.}
\end{tabular}
\label{paramtemp}
\end{table}

\subsection{Stormwater System}
\begin{figure*}[h]
\centerline{\includegraphics[trim={0.7cm 0cm 0.5cm 0.20cm},clip,width=\textwidth]{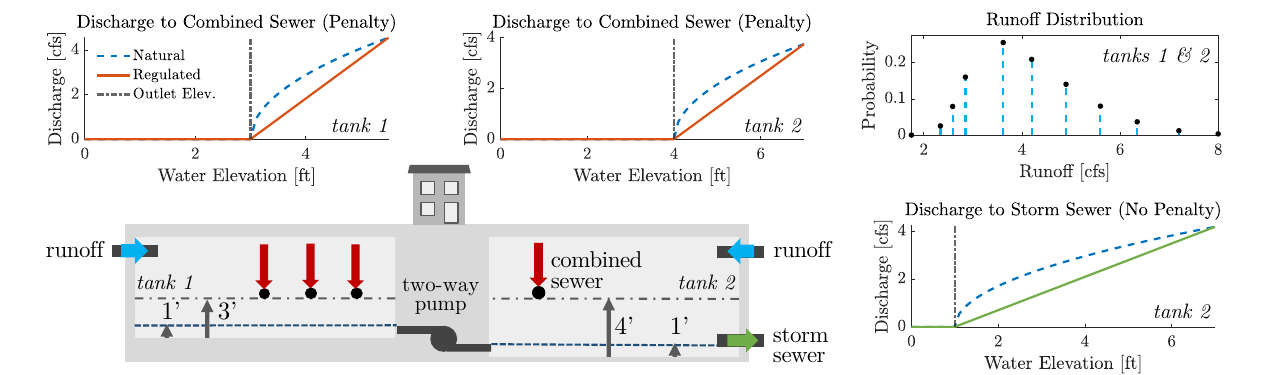}}
\caption{A stormwater system with two tanks connected by an automated pump. The pump can be operated at a maximum flow rate of 10 cfs in either direction, as long as the water level is above the 1-foot sump. When water levels are higher than the sump, tank 2 discharges stormwater passively into the storm sewer. At higher water levels, both tanks discharge stormwater passively into a combined sewer. Mechanical outlet regulators throttle discharge so it is linearly increasing with the water level. Combined sewers carry a mixture of stormwater and untreated wastewater that can overflow into natural waterways during rain events. We penalize discharge into combined sewers in our examples to demonstrate control strategies that aim to minimize the social and ecological impacts of combined sewer overflows. We present a discrete, positively skewed distribution for the random surface runoff $W_t$. The first three moments are approximately 4.0 cfs (mean), 1.2 cfs$^2$ (variance), and 0.72 (skewness).}
\label{twotanksys}
\end{figure*}

We consider a stormwater system with two tanks that are connected by an automated pump (Fig. \ref{twotanksys}). Water enters the system due to a stochastic process of surface runoff, and water exits the system by discharging to a storm sewer or, if the water level is too high, to a combined sewer. %
\emph{Combined sewers} are present in older cities (e.g., Toronto and San Francisco) and permit untreated wastewater to discharge into natural waterways, if necessary, due to limited sewer capacity. We aim to quantify and minimize the risk of combined sewer overflows, first in simulation and ultimately in practice. Large infrequent combined sewer overflows are believed to cause disproportionate ecological harm compared to smaller more frequent overflows \cite{lucas2015}. Adopting a risk-averse approach in this context may allow decision-makers to calibrate their control systems to minimize ecological harm more effectively.

We model the system as  $x_{t+1} =  x_t + F(x_t,u_t,w_t) \cdot \triangle\tau$ for $t = 0,1,\dots,N-1$
such that if $x_{t+1,i} > \overline{k}_i$, then we redefine $x_{t+1,i} := \overline{k}_i$. $x_t \in \mathbb{R}_{+}^2$ is a realization of $X_t$, and the $i^{\text{th}}$ entry $x_{it}\in [0, \overline{k}_i]$ (ft) is a water level of tank $i$. %
$u_t \in [-1,1]$ is a realization of $U_t$, representing a pump setting. $w_t \in \mathbb{R}$ (cfs) is a realization of $W_t$, representing surface runoff that enters the system during a storm. $\triangle \tau$ is the duration of $[t,t+1)$, and $N \in \mathbb{N}$ represents the duration of the storm. The function $F$ is a simplified physics-based model, $F(x,u,w)  := \left[ F_1(x,u,w), F_2(x,u,w) \right]^T \in \mathbb{R}^2$, where
\begin{equation}\begin{aligned}
F_1(x,u,w) & :=  (w - q_{\text{cso},1}(x_1) + q_\text{pump}(x,u))/a_1\\
F_2(x,u,w) & :=  (w - q_{\text{cso},2}(x_2) - q_\text{pump}(x,u) -q_\text{storm}(x_2))/a_2.
\end{aligned}\end{equation}
We define $q_{\text{cso},i}$, $q_\text{pump}$, and $q_\text{storm}$ subsequently. Please refer to Table \ref{tanksysinfo} for the model parameters. 

The combined sewer outlets are equipped with flow regulators, where each regulator produces an outflow rate $q_{\text{lin},i}$. The flow rate into the combined sewer from tank $i$, $q_{\text{cso},i}$, is expressed in terms of $q_{\text{lin},i}$ and the number of combined sewer outlets $\mathrm{N}_{\text{cs},i}$ as follows:
\begin{equation}\label{mycso}\begin{aligned}
    q_{\text{cso},i}(x_i) & := q_{\text{lin},i}(x_i) \cdot \mathrm{N}_{\text{cs},i}\\
    q_{\text{lin},i}(x_i) & := \textstyle \overline{q}_{\text{cs},i} - \frac{\overline{q}_{\text{cs},i}}{\overline{k}_i - z_{\text{cs},i}} \min\{\overline{k}_i-x_i, \overline{k}_i - z_{\text{cs},i}\},%
\end{aligned}\end{equation}
where $\overline{q}_{\text{cs},i}:= c_\text{d} \tilde{\pi} r_{\text{cs},i}^2 \big(2 \tilde{g} (\overline{k}_i - z_{\text{cs},i})\big)^{1/2}$ is tank $i$'s maximum outflow rate to the combined sewer from an outlet with radius $r_{\text{cs},i}$ and elevation $z_{\text{cs},i}$.

We define $q_\text{pump}$ so that the pumping rate is proportional to $u$ when possible, and we specify intermediary cases to permit continuity and a start-up phase. $q_\text{pump}(x,u)$ is given by 
\begin{subequations}
\begin{equation}\label{qpump1}
    q_\text{pump}(x,u) := \begin{cases}0 & \text{if }\mathcal{I}_1(x_1,u) \text{ or } \mathcal{I}_2(x_2,u) \\ \ell(x_1,u) & \text{if } x_1 \in [z_\text{p} - \epsilon,z_\text{p} + \epsilon] \text{ and } u < 0\\
    \ell(x_2,u) & \text{if } x_2 \in [z_\text{p} - \epsilon,z_\text{p} + \epsilon] \text{ and } u \geq 0\\
    u \cdot \overline{q}_{\text{p}} & \text{otherwise}\end{cases},
\end{equation}
where $\overline{q}_{\text{p}}$ (cfs) is the maximum desired pumping rate, $\ell(x_i,u)$ models a start-up phase, $z_\text{p}$ is a threshold elevation, and $\epsilon \ll z_\text{p}$ is a positive number. $\mathcal{I}_i(x_i,u)$ is true or false, depending on whether a water level is high enough for pumping,
\begin{equation}\begin{aligned}
    \mathcal{I}_1(x_1,u) & := x_1 < z_\text{p} - \epsilon \text{ and } u < 0\\
    \mathcal{I}_2(x_2,u) & :=x_2 < z_\text{p} - \epsilon \text{ and }u \geq 0.
\end{aligned}\end{equation}
For example, $\mathcal{I}_2(x_2,u)$ is true if and only if the pump attempts to push water from tank 2 to tank 1 ($u \geq 0$), but the water level in tank 2 is not high enough for pumping to occur. We define the start-up phase function by $\ell(x_i,u) := \frac{u\cdot \overline{q}_{\text{p}}}{2 \epsilon} (x_i + \epsilon - z_\text{p} )$.
\end{subequations}

The definition of $q_\text{storm}$, which resembles $q_{\text{lin},i}$ \eqref{mycso}, is $q_\text{storm}(x_2) := \overline{q}_{\text{s}} - \frac{\overline{q}_{\text{s}}}{\overline{k}_2 - z_{\text{s}}} \min\{\overline{k}_2-x_2, \overline{k}_2 - z_{\text{s}}\}$, %
where $\overline{q}_{\text{s}}:= c_\text{d} \tilde{\pi} r_{\text{s}}^2 \big(2 \tilde{g} (\overline{k}_2 - z_{s})\big)^{1/2}$ is tank 2's max outflow rate to the storm sewer from an outlet with radius $r_{\text{s}}$ and elevation $z_{\text{s}}$.

We use a discrete, positively skewed distribution for $W_t$ (Fig. \ref{twotanksys}). In previous work, we simulated a design storm in PCSWMM (Computational Hydraulics International) \cite{sustech}. PCSWMM is an extension of the US Environmental Protection Agency's Stormwater Management Model \cite{swmm}, an industry standard software package for the design of stormwater systems. We obtained samples of surface runoff from these simulations, and the empirical distribution had positive skew, which is reflected in the current distribution. %

\begin{table}[t!]
\caption{Stormwater System Parameters}
\label{table}
\setlength{\tabcolsep}{5pt}
\begin{tabular}{|p{25pt}|p{125pt}|p{50pt}|}
\hline
\textbf{Symbol} &  \textbf{Description} & \textbf{Value} \vspace{0.5mm}\\
\hline
$a_1$ & Surface area of tank 1 & 30000 ft\textsuperscript{2} \vspace{.5mm} \\
$a_2$ & Surface area of tank 2 & 10000 ft\textsuperscript{2} \vspace{.5mm} \\
$A_G$ & Control space grid &  $\{-1, 0, 1\}$ \vspace{.5mm} \\
$c_{\text{d}}$ & Discharge coefficient & 0.61  \vspace{0.5mm}\\
$ \triangle \tau $ & Duration of $[t,t+1)$ & 5 min \vspace{0.5mm}\\
$\epsilon$ & Positive number much less than pumping threshold elevation & $\frac{1}{12}$ ft \vspace{1mm}\\
$\tilde{g}$ & Acceleration due to gravity & 32.2 $\frac{\text{ft}}{\text{s}^2}$\vspace{0.5mm}\\
$k_1$ & Combined sewer (CS) outlet elevation, tank 1  & 3 ft\vspace{0.5mm}\\
$k_2$ & CS outlet elevation, tank 2  & 4 ft\vspace{0.5mm}\\
$\overline{k}_1$ & Max value of $x_1$ in state space grid & 5.5 ft\vspace{0.5mm}\\
$\overline{k}_2$ & Max value of $x_2$ in state space grid & 7 ft\vspace{0.5mm}\\
$N$ & Length of discrete time horizon & 48 ($=$ 4 h) \vspace{0.5mm}\\ 
$\mathrm{N}_{\text{cs},1}$ & Number of CS outlets in tank 1 & 3 outlets \vspace{0.5mm}\\
$\mathrm{N}_{\text{cs},2}$ & Number of CS outlets in tank 2 & 1 outlets\vspace{0.5mm}\\
$\tilde{\pi}$ & Circumference-to-diameter ratio & $\approx$ 3.14 \vspace{0.5mm}\\
$\overline{q}_\text{p}$ & Maximum desired pumping rate & 10 cfs\vspace{0.5mm}\\
$r_{\text{cs},1}$ & CS outlet radius, tank 1 & $\frac{1}{4}$ ft\vspace{1mm}\\
$r_{\text{cs},2}$ & CS outlet radius, tank 2 & $\frac{3}{8}$ ft\vspace{1mm}\\
$r_{s}$ & Storm sewer outlet radius & $\frac{1}{3}$ ft\vspace{1mm}\\
$S_G$ & State space grid & $\{0,0.1,\dots,\overline{k}_1\}\text{ ft}\times \{0,0.1,\dots,\overline{k}_2\}\text{ ft}$ \\%\vspace{0.001mm}\\ 
$z_\text{p}$ & Pumping threshold elevation & 1 ft \vspace{0.5mm}\\
$z_{\text{cs},1}$ & CS outlet elevation, tank 1 & 3 ft\vspace{0.5mm}\\
$z_{\text{cs},2}$ & CS outlet elevation, tank 2 & 4 ft \vspace{0.5mm}\\
$z_{s}$ & Storm sewer outlet elevation & 1 ft\vspace{0.5mm}\\
\hline
\multicolumn{3}{p{240pt}}{CS $=$ combined sewer, cfs $=$ cubic feet per second, ft $=$ feet, s $=$ 
seconds, min $=$ minutes, h $=$ hours.}
\end{tabular}
\label{tanksysinfo}
\end{table}

The cost function $c$ quantifies the water volume contributed to the combined sewer in hundreds of cubic feet,
\begin{equation}\label{mycostswater}
         c(x,u) = c_N(x) = q_{\text{cso}}(x) \cdot \Delta \tau \cdot 0.01.
\end{equation}
The term $q_{\text{cso}}(x)$ is the total discharge rate (cfs) to the combined sewer,
     $  \textstyle q_{\text{cso}}(x) := \sum_{i=1}^2 {q_{\text{cso},i}(x_{i})}$,
where $q_{\text{cso},i}$ is given by \eqref{mycso}. $q_{\text{cso},i}(x_{i})$ is the discharge rate (cfs) to the combined sewer from tank $i$ when the water level of tank $i$ is $x_{i}$ (ft). The definition of $q_{\text{cso}}$ assumes a constant discharge rate on each time interval $[t-1,t)$. The factor of 0.01 in \eqref{mycostswater} is used so that the resulting units are in hundreds of cubic feet. The cumulative random cost $Z$ represents the total water volume (hundreds of ft\textsuperscript{3}) that is discharged to the combined sewer during a four-hour storm.

\begin{table}[]
\centering
\caption{Computational Resources}
\begin{tabular}{l|ll|ll|}
\cline{2-5}
                                & \multicolumn{2}{c|}{\text{Thermostatic Regulator}}                    & \multicolumn{2}{c|}{\text{Stormwater System}}                      \\ \cline{2-5} 
                                & \multicolumn{1}{l|}{\textbf{Grid Size}}  & \textbf{Runtime}     & \multicolumn{1}{l|}{\textbf{Grid Size}}     & \textbf{Runtime}    \\ \hline 
\multicolumn{1}{|l|}{Problem \ref{prob1} (EU) } & \multicolumn{1}{l|}{51}      & 2 min   & \multicolumn{1}{l|}{3976}       & 10 min \\ \hline
\multicolumn{1}{|l|}{Problem \ref{cvarproblem} (CVaR)} & \multicolumn{1}{l|}{51 x 67} & 25  min & \multicolumn{1}{l|}{3976 x 491} & 136 h  \\ \hline 
\multicolumn{5}{p{240pt}}{\vspace{.5mm} To emphasize the additional complexity of Problem \ref{cvarproblem}, in the last row, we have listed the size of the computational grid as the product of the cardinality of $S_G$ and the cardinality of the discretization of $\mathcal{Z}$. Thermostatic regulator runs utilized 4 cores, and stormwater system runs utilized 30 cores in a multi-tenant cluster environment. Our analysis code is written in MATLAB (The Mathworks, Inc.) and is available from  \href{https://github.com/risk-sensitive-reachability/IEEE-TCST-2021}{https://github.com/risk-sensitive-reachability/IEEE-TCST-2021}.}
\label{runtime}
\end{tabular}
\end{table}

\section{Numerical Results: Optimal Control}\label{numresults_optimal_control}
Using the models from the previous section, we have solved Problem \ref{prob1} (EU) and Problem \ref{cvarproblem} (CVaR) numerically by implementing Algorithm \ref{valalgwhittle} and Algorithm \ref{dpcvar}, respectively.
As anticipated, \emph{significantly} reduced computational resources are required for Problem \ref{prob1} versus Problem \ref{cvarproblem}. Recall that solving Problem \ref{cvarproblem} requires an augmented state space $S \times \mathcal{Z}$, where $\mathcal{Z} := [-\bar{a},\bar{a}] \subseteq \mathbb{R}$ and $0 \leq Z' \leq \bar{a} := (\bar{d}-\underline{d})(N+1)$. Hence, the range of $\mathcal{Z}$ depends on the length of the interval $[\underline{d},\bar{d}]$ formed by the lower and upper bounds of the stage cost and the length of the time horizon $N$. The discretization of $\mathcal{Z}$ also depends on the desired precision of the cumulative costs. For example, we have chosen a precision of approximately 1 degree Celsius for the thermostatic regulator and 500 cubic feet for the stormwater system. Covering the range of $\mathcal{Z}$ at this precision increases the cardinality of the overall computational grid by 67 times and 491 times, respectively. Table \ref{runtime} outlines the resources utilized in our unoptimized implementation. We have made no attempt to improve efficiency, except for parallelizing operations in a given DP recursion.

To distinguish between an exact solution and a numerical solution returned by a computer, we introduce some notation. $\hat{V}_\theta^*$ ($\hat{J}_\alpha^*$) indicates a computation of $V_\theta^*$ ($J_\alpha^*$), and $\hat{\pi}_\theta^*$ ($\hat{\pi}^*_\alpha$) indicates a computation of a policy that is optimal for $V_\theta^*$ ($J_\alpha^*$).
Using $\hat{\pi}^*_\theta$ or $\hat{\pi}^*_\alpha$ and an initial condition $x$, we have simulated 10 million trajectories to estimate a distribution of $Z$ that is optimal w.r.t. EU or CVaR, respectively. We use these simulations to study the trends between particular empirical statistics of $Z$, which we have selected based on how EU or CVaR encode risk aversion. In the EU case, we study how the expectation $E_x^{\hat{\pi}_\theta^*}(Z)$ and the variance $\text{var}_x^{\hat{\pi}_\theta^*}(Z)$ vary with $\theta$ \eqref{approx2}.
In the CVaR case, we study how the quantile $\text{VaR}_{\alpha,x}^{\hat{\pi}^*_\alpha}(Z)$ and the expected exceedance above the quantile $E_x^{\hat{\pi}^*_\alpha}(\max\{Z - \text{VaR}_{\alpha,x}^{\hat{\pi}^*_\alpha}(Z),0\})$ vary with $\alpha$ \eqref{howencoderiskcvar}.

\begin{figure*}[!t]
\centerline{\includegraphics[trim={1cm 1cm 1.25cm 1cm},clip,width=\textwidth]{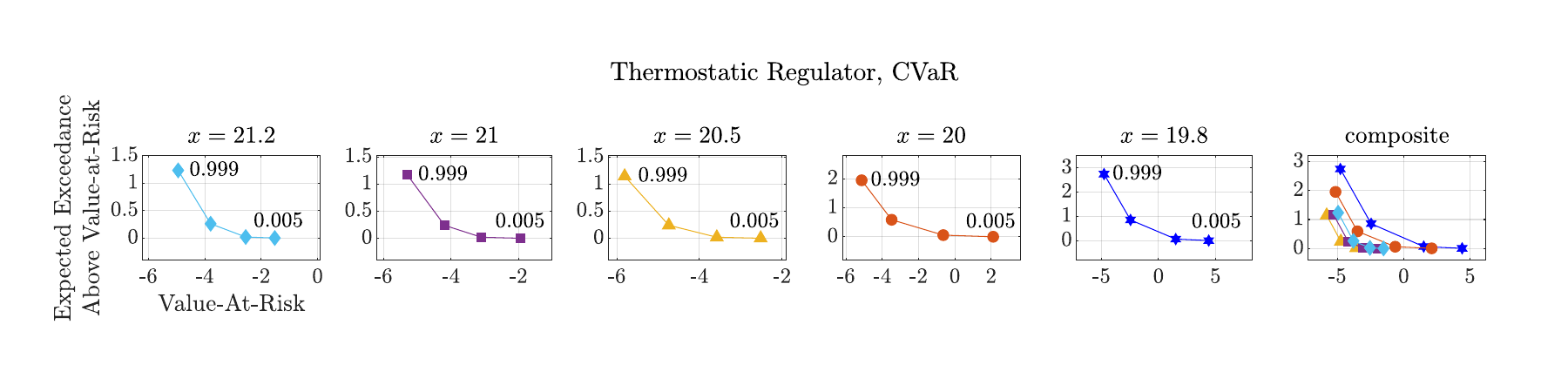}}
\caption{Empirical estimates of the trade-offs between the Value-at-Risk and the expected exceedance above the Value-at-Risk of a cumulative cost for the thermostatic regulator under CVaR-optimal control. Plots are shown for each initial condition $x \in \{19.8, 20, 20.5, 21, 21.2\}$ $^\circ$C and $\alpha \in \{0.999, 0.5, 0.05, 0.005\}$. Each point represents the result of sampling $10^{7}$ trajectories for a given $x$ and $\alpha$. According to (\ref{howencoderiskcvar}), nearly risk-neutral preferences (e.g., $\alpha = 0.999$) lead to a nearly equal weighting between the Value-at-Risk and the expected exceedance beyond the Value-at-Risk. As risk aversion increases ($\alpha$ decreases), greater emphasis is placed on minimizing the expectation of the outcomes worse than the Value-at-Risk.}
\label{temp_tradeoff_cvar}
\end{figure*}

\begin{figure}
\begin{subfigure}[t!]{\columnwidth}

\includegraphics[trim={0.4cm 0 1.1cm 0},clip,width=\columnwidth]{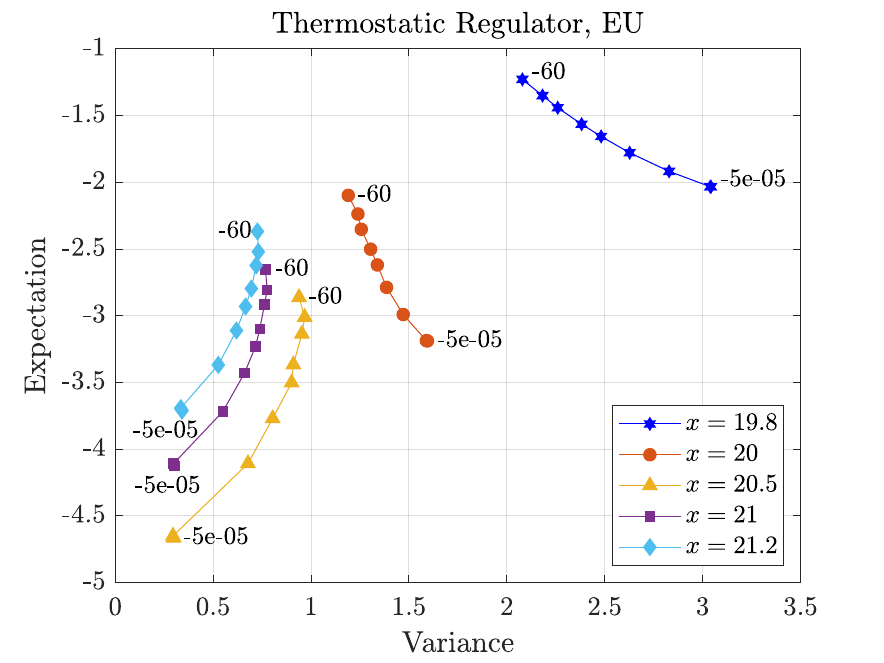}
\end{subfigure}
\begin{subfigure}[t!]{\columnwidth}
\includegraphics[trim={8.25cm 2cm 8cm 2cm},clip,width=\columnwidth]{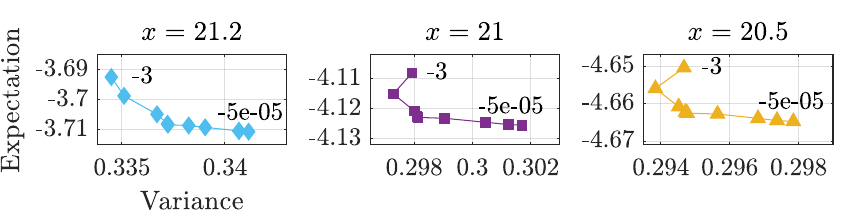}
\end{subfigure}

\caption{Estimates of $E_x^{\hat{\pi}_\theta^*}(Z)$ versus $\text{var}_x^{\hat{\pi}_\theta^*}(Z)$ for the thermostatic regulator under EU-optimal control after $10^{7}$ samples for each $x$ and $\theta.$  Plots are shown for each initial condition $x \in \{19.8, 20, 20.5, 21, 21.2\}$ $^\circ$C, and $\theta \in \{-5\mathrm{e}{-5}, -3, -9, -12, -15, -18, -24, -30, -60\}$ (\textbf{top}). In the top plot, the points that correspond to $\theta \in \{-5\mathrm{e}{-5}, -3\}$ overlap for each $x$. There is a neighborhood $\mathcal{N}_x$ of zero in which making $\theta$ more negative leads to a reduction in the variance at the expense of an increase in the mean. For $x = 19.8$ and $x = 20$, the policy $\hat{\pi}_\theta^*$ with $\theta = -60$ leads to a distribution with a small variance at the expense of having a large expectation. For $x = 20.5$, $x = 21$, and $x=21.2$, the policy $\hat{\pi}_\theta^*$ with $\theta = -60$ yields a distribution with a large expectation and a large variance. However, these latter initial conditions achieve a trivial mean-variance trade-off if $|\theta|$ is sufficiently small (\textbf{bottom}). The bottom plots show $\theta \in \{-5\mathrm{e}{-5}, -0.5, -0.75, -1, -1.25, -1.5, -1.75, -2, -2.5, -3\}$. These plots indicate that if $x = 20.5$ or $x = 21$, $\mathcal{N}_x$ is approximately $(-2.5,0)$; if $x = 21.2$, $\mathcal{N}_x$ is roughly $(-3,0)$.}

\label{eu_tradeoff_temp}
\end{figure}

\subsubsection{Thermostatic Regulator} 
We present results for the thermostatic regulator in which the disturbance has a right-skewed distribution (Fig. \ref{temperature_disturbance}). First, we consider the CVaR setting. By (\ref{howencoderiskcvar}), minimizing the CVaR of $Z$ at level $\alpha \in (0,1)$ is equivalent to minimizing a linear combination of the Value-at-Risk at level $\alpha$, $\text{VaR}_{\alpha,x}^\pi(Z)$, %
and the expected exceedance above the $\text{VaR}_{\alpha,x}^\pi(Z)$, which is $E_x^{\pi}(\max\{Z - \text{VaR}_{\alpha,x}^{\pi}(Z),0\})$. %
The Value-at-Risk at level $\alpha$ is a quantile that represents the best outcome among the $\alpha \cdot 100\%$ of the worst outcomes. %
All else being equal, it is desirable to keep the Value-at-Risk for a given $\alpha$ as small as possible. However, the Value-at-Risk is not sensitive to the distribution of the values that exceed it. CVaR compensates for this limitation by incorporating a measure of the expected exceedance beyond the Value-at-Risk. We provide empirical estimates of the Pareto-efficient trade-offs between these two quantities with $\pi =\hat{\pi}^*_\alpha$, that is,  $E_x^{\hat{\pi}^*_\alpha}(\max\{Z - \text{VaR}_{\alpha,x}^{\hat{\pi}^*_\alpha}(Z),0\})$ versus $\text{VaR}_{\alpha,x}^{\hat{\pi}^*_\alpha}(Z)$, for the thermostatic regulator in Fig. \ref{temp_tradeoff_cvar}.

Now, we consider the EU setting, in which the mean-variance trend is of interest.
When the mean-variance approximation \eqref{approx2} is valid, 
the magnitude of $\theta$ represents the amount a decision-maker is willing to increase the expectation for a unit reduction in the variance. Therefore, making $\theta$ more negative should
prioritize a reduction in the variance $\text{var}_x^{\pi}(Z)$ rather than the expectation $E_x^{\pi}(Z)$. We show plots of the empirical estimates of the mean and variance for the thermostatic regulator under EU-optimal control in Fig. \ref{eu_tradeoff_temp}.

For each initial condition $x$, there is a small neighborhood $\mathcal{N}_x$ of zero such that varying $\theta$ in $\mathcal{N}_x$ leads to a mean-variance trade-off. The size of this neighborhood depends on the initial condition. The top portion of Fig. \ref{eu_tradeoff_temp} shows that if $x = 19.8$ or $x = 20$ ($^\circ$C), the variance is reduced at the expense of the mean, as $\theta$ decreases from $-5\mathrm{e}{-5}$ to $-60$. However, the bottom portion of Fig. \ref{eu_tradeoff_temp} shows that this trend only exists for a narrow range of $\theta$ if $x = 20.5$, $x = 21$, or $x=21.2$. For these latter initial conditions, varying $\theta$ in $\mathcal{N}_x$ leads to a trade-off that is practically trivial, and increasing the magnitude of $\theta$ outside of $\mathcal{N}_x$ leads to increases in both the mean and variance (see Fig. \ref{eu_tradeoff_temp}, top).

\begin{figure}[t!]
\centerline{\includegraphics[trim={0.5cm 0 1cm 0},clip,width=\columnwidth]{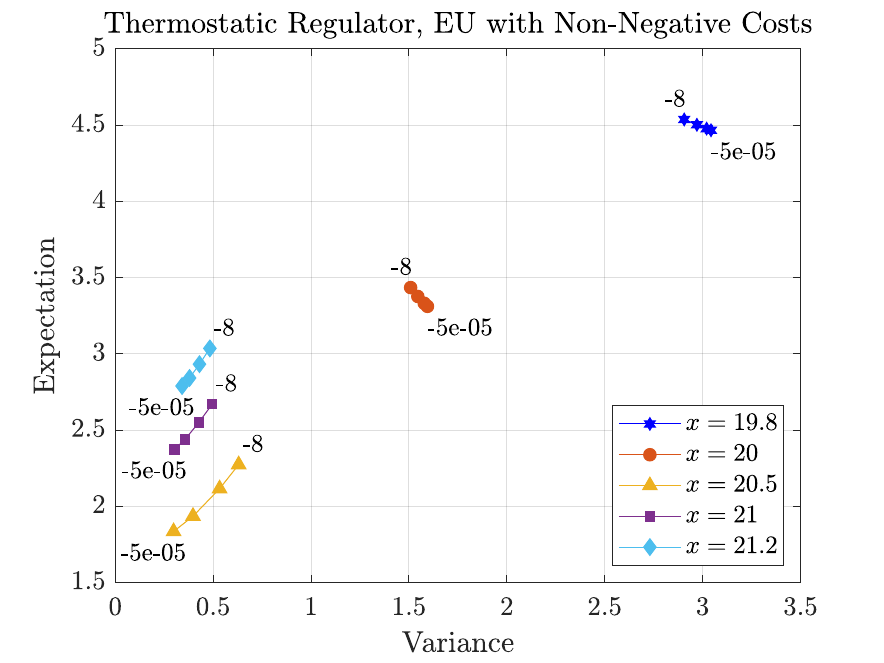}}
\caption{Estimates of $E_x^{\hat{\pi}_\theta^*}(Z')$ versus $\text{var}_x^{\hat{\pi}_\theta^*}(Z')$ for the thermostatic regulator, where we have implemented Alg. \ref{valalgwhittle} with the restriction to non-negative costs (Remark \ref{remarkmod}). These empirical moments were estimated from $10^{7}$ samples for each $x \in \{19.8, 20, 20.5, 21, 21.2\}$ $^\circ$C and $\theta \in \{-5\mathrm{e}{-5}, -6, -7, -8\}.$}

\label{nonnegative_temp_exponential_utility}
\end{figure}

\begin{figure}[t!]
\centerline{\includegraphics[trim={.65cm 0 0.65cm 0},clip,width=\columnwidth]{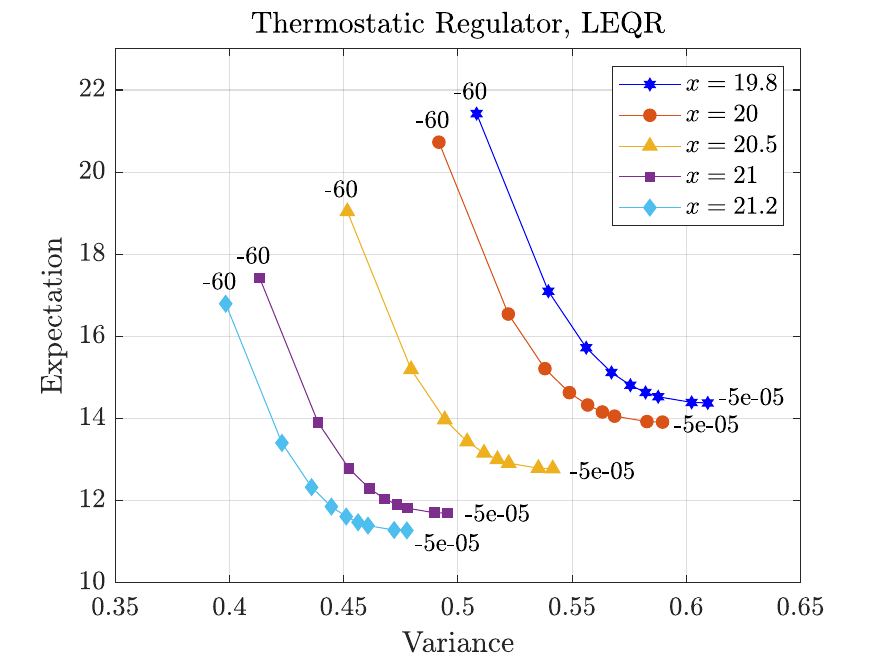}}
\caption{Empirical mean-variance curves for the thermostatic regulator under a LEQG controller \cite[Thm. 3]{whittle1990risk}, $x \in \{19.8, 20, 20.5, 21, 21.2\}$ $^\circ$C, and $\theta \in \{-5\mathrm{e}{-5}, -12, -30, -35, -40, -45, -50, -55, -60\}$. Each point represents the outcome of $10^7$ sampled trajectories for a given $x$ and $\theta$. The state vector is $\tilde x_t := x_t - b$, and the dynamics equation is $\tilde x_{t+1} = A \tilde x_t + Bu_t + w_t$, where $A = a$ and $B = (a-1)\eta R P$ (Table \ref{paramtemp}). The disturbance process is i.i.d. zero-mean Gaussian noise with variance $\sigma^2 = 0.03$. The initial condition is a Gaussian random variable with mean $x - b$ and variance $\sigma^2/100$. The random cost $Z$ is quadratic, and a realization of $Z$ takes the form, $z =  \tilde q \tilde x_N^2 + \sum_{t=0}^{N-1}  \tilde q \tilde x_t^2 +  r u_t^2$, where $\tilde q= 0.01$ and $r = 1$.}
\label{qtgstemperaturesystem}
\end{figure}

We find that similar trends occur, see Fig. \ref{nonnegative_temp_exponential_utility}, when we implement Algorithm \ref{valalgwhittle} with non-negative costs (Remark \ref{remarkmod}). Here, the most negative value of $\theta$ is $-8$ because the algorithm suffers from numerical instabilities for more negative values of $\theta$.
Notably, in the classical LEQG setting, we find consistent mean-variance trade-offs for a wide range of $\theta$ (Fig. \ref{qtgstemperaturesystem}, see caption for details regarding the simulation setting). The classical LEQG controller is linear state feedback, and a Riccati recursion provides the optimal control gains \cite[Thm. 3]{whittle1990risk}. The recursion is well-defined for negative values of $\theta$ that satisfy a condition that depends on the Riccati matrices and the noise covariance \cite[Thm. 3]{whittle1990risk}.

\subsubsection{Stormwater system}

\begin{figure*}[!t]
\centerline{\includegraphics[trim={1.2cm 1.4cm 1.2cm 1.2cm},clip,width=\textwidth]{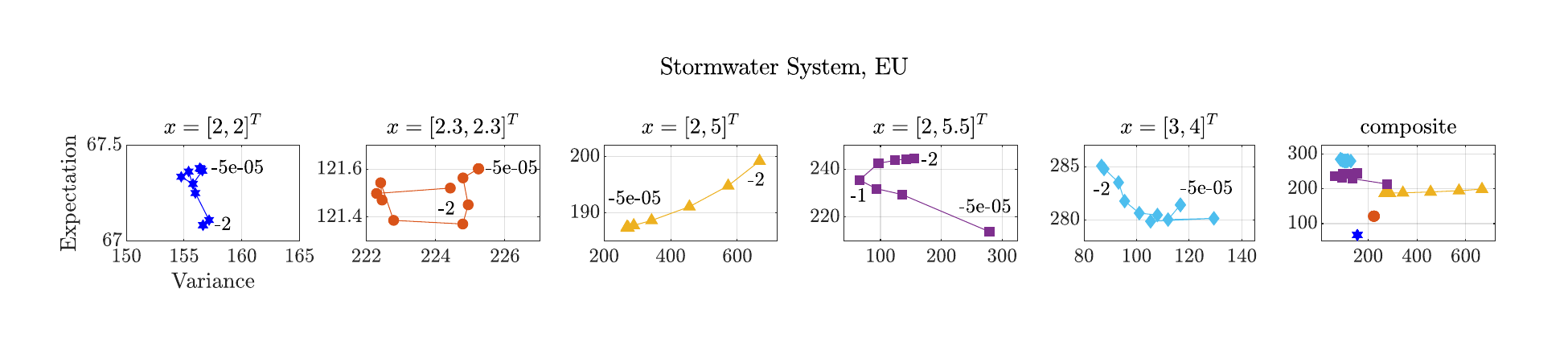}}
\caption{These plots show estimates of $E_x^{\hat{\pi}_\theta^*}(Z)$ versus $\text{var}_x^{\hat{\pi}_\theta^*}(Z)$ for the stormwater system under EU-optimal control. $Z$ is a non-negative cumulative cost, expressed in hundreds of ft$^3$ of water. Each point represents the result of sampling $10^{7}$ trajectories for a given $x$ and $\theta$. For this system, we have found the numerically stable range for $\theta$ to be approximately $-5\mathrm{e}{-5}$ to $-2$, and the plots show $\theta \in \{-5\mathrm{e}{-5}, -5\mathrm{e}{-4}, -0.005, -0.05, -0.5, -1, -1.25, -1.5, -1.75, -2\}$ for select initial water levels. For $x=[2, 2]^T$ and $x=[2.3, 2.3]^T$, no consistent mean-variance trends are present, and from the perspective of the composite plot (far right), one may consider the changes in the mean and variance to be negligible. %
If the system starts from $x=[2, 5]^T$, the mean and variance increase as $\theta$ becomes more negative. If $x=[2, 5.5]^T$, there is a mean-variance trade-off if $|\theta|$ is sufficiently small, but the mean and variance increase as $\theta$ decreases from $-1$ to $-2$. %
Trajectories starting from $x=[3, 4]^T$ exhibit a mean-variance trade-off as $\theta$ varies across some, but not all, sub-intervals of $[-2,0)$.}
\vspace{-3mm}
\label{water_tradeoff_eu}
\end{figure*}

\begin{figure}[t]
\centerline{\includegraphics[trim={0.4cm 0 0.8cm 0},clip,width=\columnwidth]{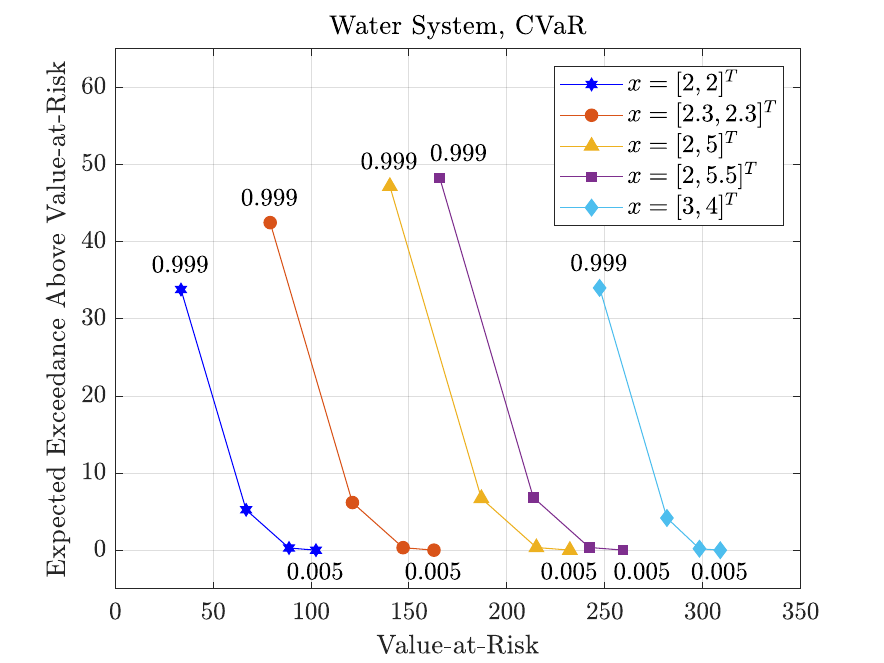}}
\caption{These plots provide estimates of the trade-offs between the Value-at-Risk $\text{VaR}_{\alpha,x}^{\hat{\pi}^*_\alpha}(Z)$ and the expected exceedance above the Value-at-Risk $E_x^{\hat{\pi}^*_\alpha}(\max\{Z - \text{VaR}_{\alpha,x}^{\hat{\pi}^*_\alpha}(Z),0\})$ of a cumulative cost $Z$ for the stormwater system under CVaR-optimal control. Plots are shown for different initial conditions $x$ and risk-aversion levels $\alpha \in \{0.999, 0.5, 0.05, 0.005\}$. Each point represents the result of sampling $10^{7}$ trajectories for a given $x$ and $\alpha$.}
\label{water_tradeoff_cvar}
\end{figure}

When EU-optimal control is applied to the stormwater system, we find that the mean-variance interpretation is again dependent on the initial conditions. In Fig. \ref{water_tradeoff_eu}, we show the empirical mean-variance curves from select initial water levels $x \in \mathbb{R}_+^2$ and values of $\theta$. The above examples demonstrate that special care must be taken when using EU-optimal control. It is known that the mean-variance approximation for EU, $\rho_{\theta,x}^\pi(Z) \approx E_x^\pi(Z) - \textstyle\frac{\theta}{4}\text{var}_x^\pi(Z)$ \eqref{approx2}, is valid under a restricted set of conditions. One of our contributions is to demonstrate that such theoretical requirements cannot be ignored in practical applications of EU-optimal control. Indeed, for a given system model and initial condition, the size of $\mathcal{N}_x$ is not known \emph{a priori} and may be trivial in practice. More importantly, we have shown that inadvertently making $\theta$ too negative can lead to a distribution with a higher mean and a higher variance.

On the other hand, CVaR-optimal control provides a consistent trade-off between the Value-at-Risk, $\text{VaR}_{\alpha,x}^{\hat{\pi}^*_\alpha}(Z)$, and the expected exceedance above the Value-at-Risk, $E_x^{\hat{\pi}^*_\alpha}(\max\{Z - \text{VaR}_{\alpha,x}^{\hat{\pi}^*_\alpha}(Z),0\})$, as $\alpha$ becomes more risk averse (closer to zero). Intuitively, this trade-off is not surprising because for any random variable $Y$ with a fixed distribution, $E(\max\{Y - \text{VaR}_{\alpha}(Y),0\})$ versus $\text{VaR}_{\alpha}(Y)$ forms a non-increasing trend, as $\alpha$ becomes more risk averse. While related, the setting of CVaR-optimal control is distinct because the distribution of $Z$, $P_x^{\hat{\pi}^*_\alpha}$, varies with $\alpha$. %
The consistent trade-off that arises from CVaR-optimal control is useful for analyzing the performance of a control system with respect to competing objectives and varying degrees of pessimism. Fig. \ref{water_tradeoff_cvar} shows a summary of performance for one particular design. However, such curves could be estimated for multiple candidate designs, and overlaying these curves could provide a concise visual comparison of performance. The current computational requirements of CVaR-optimal control restrict such comparisons to systems with low-dimensional models.

Thus far, we have analyzed system behavior and the distribution of outcomes under such behavior from select initial conditions. 
In addition, it may be useful to assess the performance of a system on its state space as a whole. An approach for this task is to use the notion of a safe set, which is the topic of the following section.

\section{Safety Analysis via Exponential Utility and CVaR: Interpretations and Examples}\label{SecIV}
One can assess the performance of a control system by studying the level sets of an optimal value function. The level sets may be called safe sets, reachable sets, or invariant sets, for example, where the precise name is chosen according to the value function's interpretation. Developing algorithms to compute these sets and the associated optimal policies is the core aim of Hamilton-Jacobi reachability analysis \cite{mitchell2005time} \cite{chen2018hamilton}, minimax safety analysis \cite{bertsekas1971minimax}, and stochastic safety analysis \cite{abate2008probabilistic, summers2010verification, ding2013, yang2018dynamic} (recall our introduction).
By formulating and solving robust or stochastic optimal control problems, these methods provide principled alternatives to Monte Carlo simulation to synthesize policies and assess whether a system can operate well in an uncertain environment. %

Here, we examine the use of risk-averse optimal control for this assessment. %
We define \emph{Exponential-Utility-safe sets} in terms of $V_\theta^*$, $\mathcal{E}_\theta^r := \{ x \in S : V_\theta^*(x)  \leq r \}$,
and \emph{CVaR-safe sets} in terms of $J_\alpha^*$, $\mathcal{C}_\alpha^r := \{ x \in S : J_\alpha^*(x)  \leq r \}$, where $r \in \mathbb{R}$.
We use the term safe sets because $V^*_\theta$ and $J_\alpha^*$ represent an optimal deviation between the state trajectory and a desired operating region. An EU-safe set $\mathcal{E}_\theta^r$ is the set of initial conditions $x \in S$ from which $V_\theta^*(x)$ is no more than $r$. If a decision-maker has a fixed price $\theta$ they are willing to pay to reduce variance and if \eqref{approx2} is valid, then $V_\theta^*(x)$ is the optimal certainty equivalent. 
The CVaR-safe set $\mathcal{C}_\alpha^r$ represents the set of initial conditions from which the optimal expected value of $Z$ in the $\alpha \cdot 100$\% worst cases is no more than $r$. This interpretation is exact if $Z$ is a continuous random variable for all $x \in S$ and $\pi \in \Pi$, for example.

\begin{figure}[!t]
\begin{subfigure}[t!]{\columnwidth}
\includegraphics[trim={1cm 0cm 1cm 0cm},clip,width=\columnwidth]{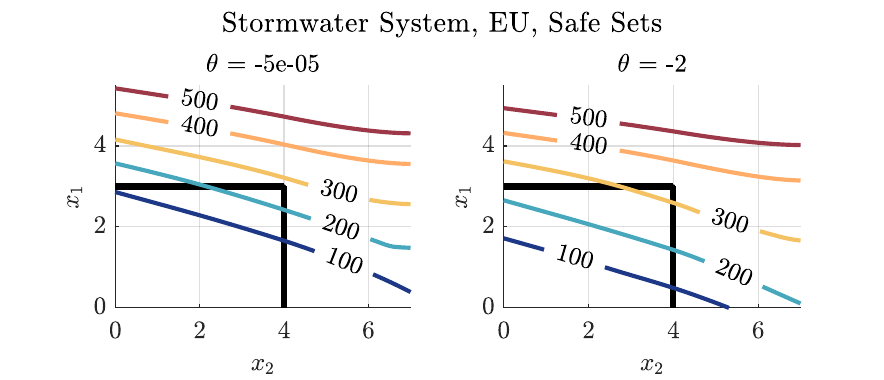}
\end{subfigure}
\begin{subfigure}[t!]{\columnwidth}
\includegraphics[trim={1cm 0cm 1cm 0cm},clip,width=\columnwidth]{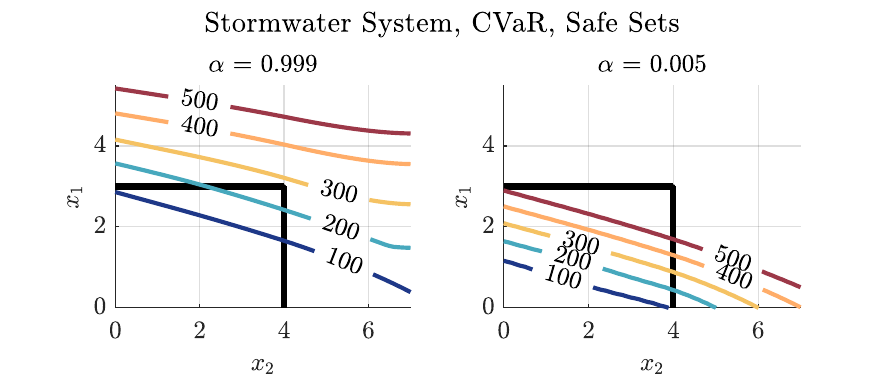}
\end{subfigure}
\caption{Estimates of EU-safe sets $\mathcal{E}_\theta^r := \{ x \in S : V_\theta^*(x)  \leq r \}$ (\textbf{top}) and CVaR-safe sets $\mathcal{C}_\alpha^r := \{ x \in S : J_\alpha^*(x)  \leq r \}$ (\textbf{bottom}) for the stormwater system, where $x = [x_1,x_2]^T \in S$.} Water levels outside of $K$ (shown in black) cause discharges to a combined sewer. We show $r \in \{100,200,\dots,500\}$ (hundreds of ft\textsuperscript{3} of water discharged to a combined sewer) for a nearly risk-neutral setting ($\theta$ near zero, $\alpha$ near 1) and a risk-averse setting ($\theta = -2$, $\alpha = 0.005$).
\vspace{-3mm}
\label{safesets}
\end{figure}

EU-safe sets and CVaR-safe sets for the stormwater system are shown in Fig. \ref{safesets}. As anticipated, the safe sets in the nearly risk-neutral setting ($\theta$ near zero, $\alpha$ near one) are indistinguishable (Fig. \ref{safesets}, left). The EU-safe sets with $\theta = -2$, the most negative value that is numerically stable, are noticeably larger than the CVaR-safe sets with $\alpha = 0.005$ (Fig. \ref{safesets}, right). It is easier to observe the contraction in the contours of the CVaR-safe sets as $\alpha$ becomes more risk averse, suggesting that CVaR may be more suitable for visually conveying the effects of varying degrees of risk aversion. 

\section{Conclusions}\label{SecV}
In this paper, we have studied the use of Exponential Utility (EU) and CVaR as risk-averse performance criteria for control systems. The optimization of EU is considerably simpler in theory and in practice. However, we have demonstrated that a non-trivial mean-variance trade-off need not occur, and making $\theta$ more negative can yield a higher variance and a higher mean. Therefore, EU-optimal control must be used cautiously. In addition to the above concerns, the choice of a more appropriate risk-averse functional may depend on several factors, which we highlight below.

\emph{Utility functions}. EU-optimal control assumes a utility function of the form $\nu_\theta(y) := e^{\frac{-\theta}{2}y}$. %
In contrast, CVaR-optimal control does not require a utility function.
There are many utility functions available, and it may be difficult to choose one that describes the desired preferences for all possible outcomes for a particular application. Some utility functions transform the costs in ways that are inappropriate for the application. 

\emph{Parameter interpretations}.
The parameter $\theta$ of EU represents an exponential aversion to larger values of $Z'$ in general \eqref{exputility} and a linear aversion to the variance when \eqref{approx2} is valid. In contrast, the risk-sensitivity level $\alpha$ that parametrizes CVaR corresponds to a fraction of the largest values of $Z$ \eqref{cvardef}.
Parameters with intuitive and precise interpretations, such as $\alpha$, may be particularly useful for applications that require the development and satisfaction of safety or performance specifications.

\emph{Risk interpretations}. 
Recall that the EU of $Z$ encodes risk using a subjective utility function $\nu_\theta$, whereas the CVaR of $Z$ encodes risk in terms of the expected amount that $Z$ exceeds a quantile \eqref{howencoderiskcvar}. In EU, the utility function is applied to all possible outcomes, while CVaR focuses exclusively on outcomes in the upper tail. 
We have provided numerical examples of a stormwater system and a thermostatic regulator to highlight how different ways to quantify risk may be more suitable for different applications.

Regulations for stormwater systems often specify design criteria in terms of quantiles, such as adequate performance under the 10-percentile (10-year) and 1-percentile (100-year) storm events. In these assessments, standard practices measure performance in terms of a random volume of overflow ($Z$) \cite{usepa2016}. CVaR may be more appropriate than EU in these circumstances, because it encodes risk in terms of quantiles and does not distort the units of $Z$ (a physical quantity) through a utility function. However, the computational resources that are required for CVaR-optimal control are not economically feasible for higher-dimensional stormwater systems in practice. While EU-optimal control has the benefit of significantly reduced computational requirements, for the previously mentioned reasons, it is unlikely to be useful for stormwater management, or more broadly, for applications with objective performance criteria.

EU-optimal control is better suited for applications with subjective performance criteria.
For example, it is natural to measure the performance of indoor heating and cooling systems in terms of perceived comfort. %
$Z$ may reflect a random deviation from the desired air temperature, and extreme values of $Z$ may be inconvenient but not safety-critical. Office buildings can simply close during rare catastrophic cooling failures. To maximize comfort in normal circumstances, $Z$ should be small on average and have low variance \cite{humphreys1979}. Moreover, occupant satisfaction with heating and cooling systems is related to the level of perceived control over the indoor environment \cite{hellwig2015}. In addition to its computational advantages, EU may be more appropriate than CVaR in this setting due to its connection to mean-variance control for values of $\theta$ in a neighborhood of zero. Unfortunately, our simple thermostatic regulator example has cast doubt on whether $\theta$ provides satisfactory mean-variance control in practice. 

While risk-averse optimal control is theoretically attractive in principle, the limitations of current approaches inhibit their adoption in practice. 
The gap between theory and practice motivates several exciting avenues for future investigations. In particular, we see value in developing:
\begin{enumerate}
    \item new methods to identify the classes of control systems in which EU provides non-trivial mean-variance trade-offs,
    \item scalable approximations for CVaR-optimal control by leveraging stochastic rollout or other grid-free policy improvement techniques, and
    \item efficient strategies for assessing the degree to which a given system could benefit from a risk-averse approach relative to a conventional one.
\end{enumerate}
Advances in these areas are needed to fully develop the potential advantages of risk-averse control as a decision-making framework that can accommodate competing objectives and varying degrees of pessimism about an uncertain future.

\section*{Acknowledgments}
The authors thank Laurent Lessard, Claire Tomlin, Marco Pavone, Chuanning Wei, and Yuxi Han for fruitful discussions.
\bibliographystyle{ieeecolor}

\newpage
%\onecolumn
\begin{IEEEbiography}[{\includegraphics[width=1in,height=1.25in,clip,keepaspectratio]{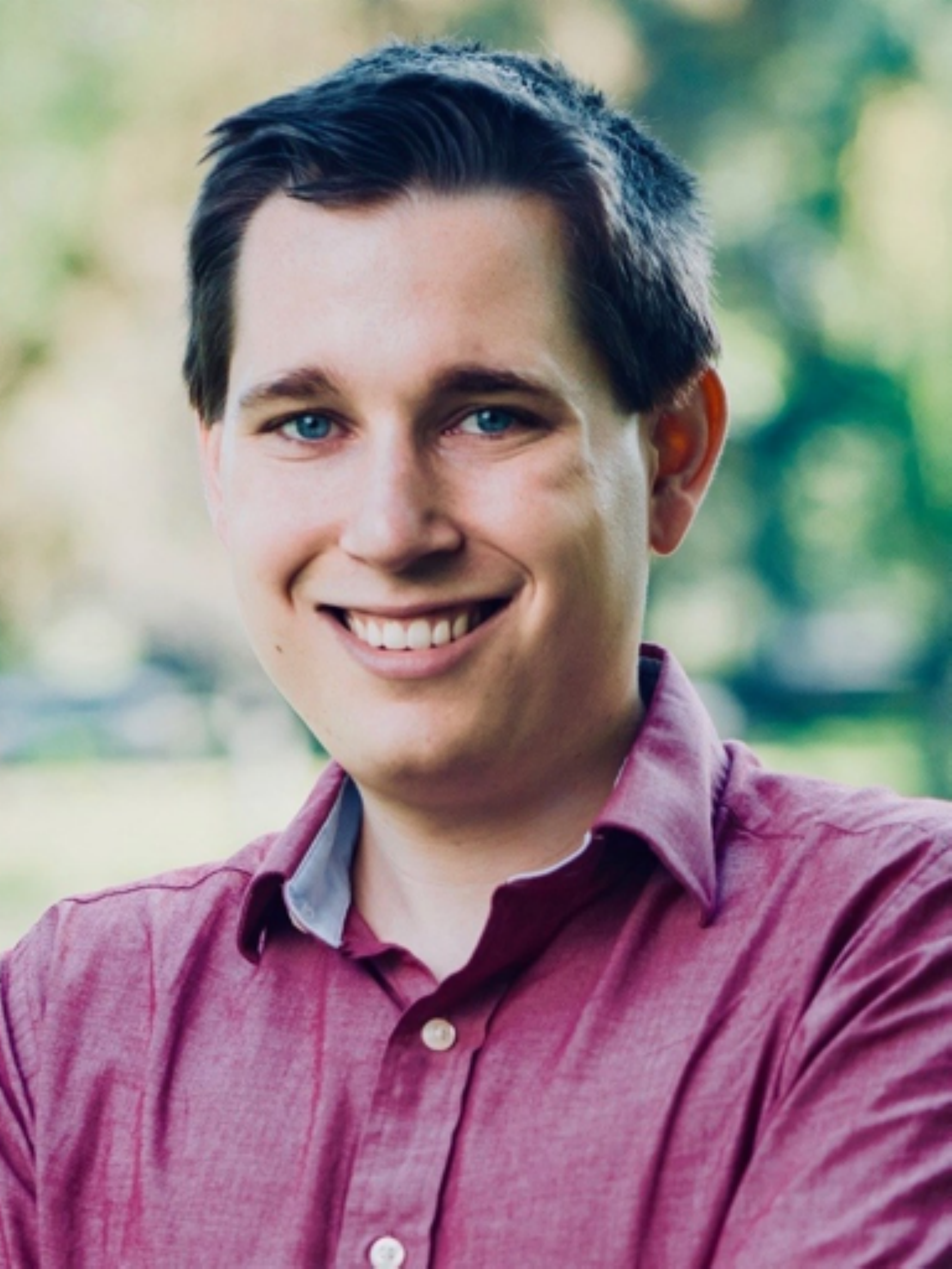}}]{Kevin Smith} is a Ph.D. candidate in Environmental and Water Resources Engineering at Tufts University and a recipient of the NSF Integrative Graduate Education and Research Traineeship (IGERT) on Water and Diplomacy and the NSF Research Traineeship (NRT) on Data Driven Decision Making to Address Complex Resource Problems. Kevin is also Director of Product at OptiRTC, Inc., where he is responsible for developing flexible real-time systems for the continuous monitoring and adaptive control of stormwater infrastructure. Kevin earned his B.A. in Environmental Studies from Oberlin College and his B.S. in Earth and Environmental Engineering from Columbia University. Kevin's research seeks to understand the opportunities and risks associated with semi-autonomous civil infrastructure when considered as a technology for mediating environmental conflicts.
\end{IEEEbiography}
\vspace{-150mm}
\begin{IEEEbiography}[{\includegraphics[width=1in,height=1.25in,clip,keepaspectratio]{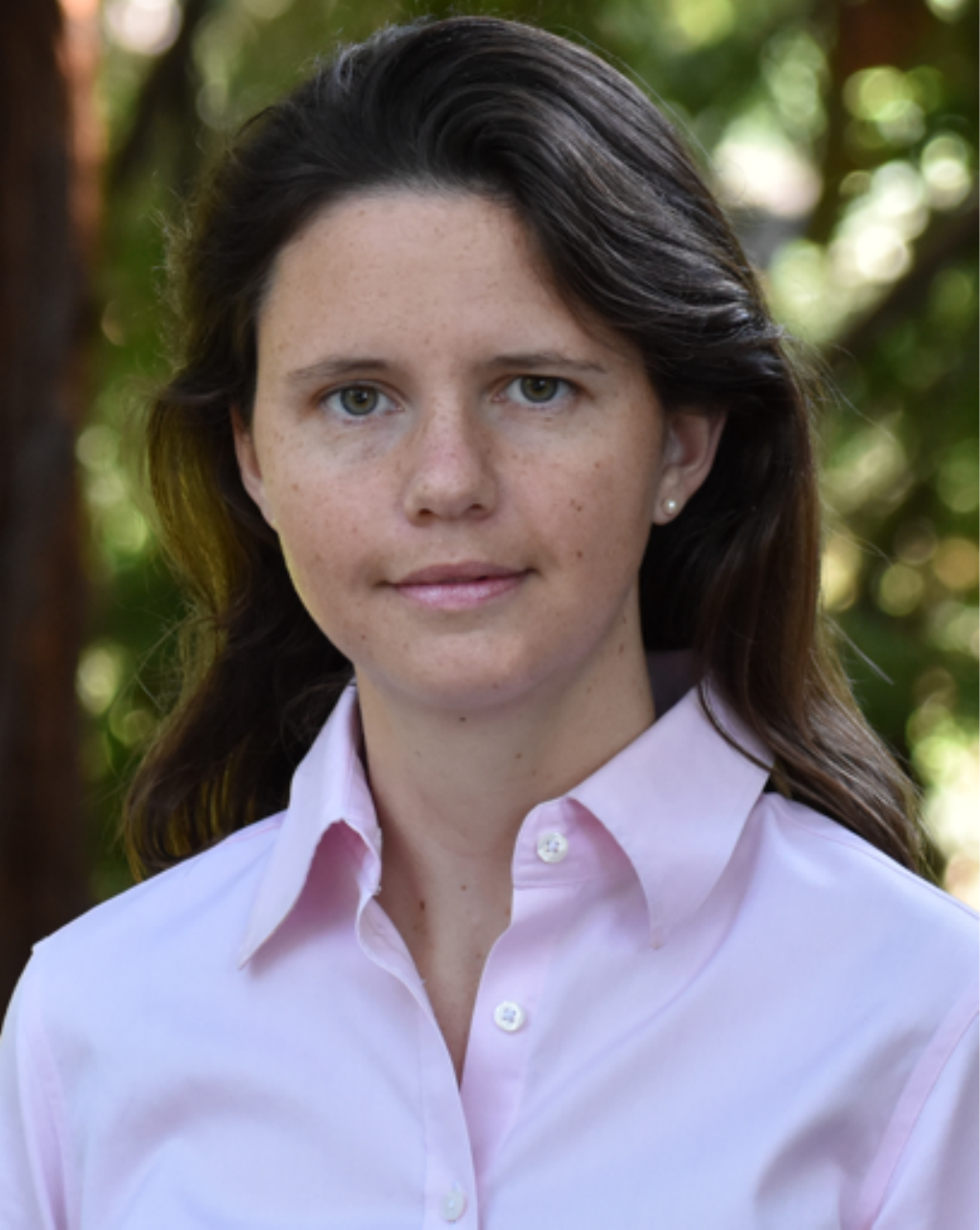}}]{Margaret Chapman} is an Assistant Professor with the Electrical and Computer Engineering Department, University of Toronto, Toronto, Canada. Her research focuses
on risk-sensitive and stochastic control, with emphasis on safety analysis and applications in
healthcare and sustainable cities. Margaret is a recipient of the 2021 Leon O. Chua Award for achievement in nonlinear science (Electrical Engineering and Computer Sciences, UC Berkeley). She has also received a US National Science Foundation Graduate Research Fellowship, a Berkeley Fellowship for Graduate Study, and a Stanford University Terman Engineering Scholastic Award. \end{IEEEbiography}

\newpage
\onecolumn
\section*{Supplementary Material}

%supp_material.tex
\noindent This document provides some technical details to accompany the main paper. %The reference numbers in the current document pertain to the works listed at the end of the document.

\section{Notes Regarding Exponential Utility}
\subsection{Note Regarding Limit}
Recall the statement from the main paper: it can be shown under certain conditions that $\underset{\theta \rightarrow 0}{\lim} \rho_{\theta,x}^\pi(Z) = \underline{b} + E_x^\pi(Z')$. We explain this statement by providing conditions under which  $\underset{\theta \rightarrow 0}{\lim} \rho_{\theta,x}^\pi(Z') = E_x^\pi(Z')$ holds.\\\\
Assume that $Z' \in L^2 := L^2(\Omega,\mathcal{B}_\Omega,P_x^\pi)$ and there are real numbers $a$ and $b$ such that $a < 0 < b$ and $\exp(\frac{-\theta}{2}Z') \in L^2$ for all $\theta \in [a,b]$. Under these conditions, one can use \cite[Thm. 2.27, p. 56]{folland1999real} and H\"{o}lder's Inequality to find that
\begin{align}
  \textstyle   \frac{\mathrm{d}}{\mathrm{d}\theta}E_x^\pi(\exp(\frac{-\theta}{2}Z')) & = \textstyle \frac{-1}{2} E_x^\pi(Z' \exp(\frac{-\theta}{2}Z'))\label{my1} \\
  \textstyle   \underset{\theta \rightarrow 0}{\lim} E_x^\pi(Z' \exp(\frac{-\theta}{2}Z')) & = E_x^\pi(Z')\label{my22}\\
   \textstyle   \underset{\theta \rightarrow 0}{\lim} E_x^\pi(\exp(\frac{-\theta}{2}Z')) & = 1. \label{my33}
\end{align}
%$\frac{\mathrm{d}}{\mathrm{d}\theta}E_x^\pi(\exp(\frac{-\theta}{2}Z')) = \frac{-1}{2} E_x^\pi(Z' \exp(\frac{-\theta}{2}Z'))$, $\underset{\theta \rightarrow 0}{\lim} E_x^\pi(Z' \exp(\frac{-\theta}{2}Z')) = E_x^\pi(Z')$, and $\underset{\theta \rightarrow 0}{\lim} E_x^\pi(\exp(\frac{-\theta}{2}Z')) = 1$. 
%
For \eqref{my1}, define $\tilde f : \Omega \times [a,b] \rightarrow \mathbb{R}$ such that $\tilde f(\omega, \theta) := \exp(\frac{-\theta}{2}Z'(\omega))$. It holds that $\tilde f(\cdot,\theta) \in L^2 \subseteq L^1 := L^1(\Omega,\mathcal{B}_\Omega,P_x^\pi)$ for all $\theta \in [a,b]$. The partial derivative of $\tilde f$ with respect to $\theta$ is given by
\begin{equation}
\textstyle \frac{\partial}{\partial \theta}\tilde f(\omega, \theta) = \frac{-1}{2} Z'(\omega) \exp(\frac{-\theta}{2}Z'(\omega)),
\end{equation}
and
\begin{equation}\label{my55}
 \textstyle   |\frac{\partial}{\partial \theta}\tilde f(\omega, \theta)| \leq \frac{1}{2} Z'(\omega) \exp(\frac{|a|}{2}Z'(\omega)) \;\;\; \forall \omega \in \Omega \; \; \forall \theta \in [a,b].
\end{equation}
We denote the function on the right of \eqref{my55} by $\tilde g := \frac{1}{2} Z'\exp(\frac{|a|}{2}Z')$. To derive \eqref{my55}, note that $a < 0$ and 
\begin{equation}
    -b \leq -\theta \leq -a = |a| \;\;\; \;\;\forall \theta \in [a,b].
\end{equation}
By H\"{o}lder's Inequality \cite[p. 82]{ash1972probability}, we know that $\tilde g \in L^1$ because $Z' \in L^2$ and $\exp(\frac{-\theta}{2}Z') \in L^2$ for all $\theta \in [a,b]$, and in particular, for $\theta = a$. Then, we use \cite[Thm. 2.27b]{folland1999real}, which allows us to interchange the derivative and the integral, to conclude \eqref{my1}.\\\\
To show \eqref{my22}, define $\bar{f} : \Omega \times [a,b] \rightarrow \mathbb{R}$ such that $\bar{f}(\omega,\theta) := Z'(\omega) \exp(\frac{-\theta}{2}Z'(\omega))$. Note that $\bar{f}(\cdot,\theta) \in L^1$ for all $\theta \in [a,b]$ as a consequence of H\"{o}lder's Inequality. It holds that
\begin{equation}
 \textstyle   |\bar{f}(\omega,\theta) | \leq \bar{g}(\omega) := Z'(\omega) \exp(\frac{|a|}{2}Z'(\omega)) \;\;\; \forall \omega \in \Omega \; \; \forall \theta \in [a,b],
\end{equation}
where $\bar{g} \in L^1$ by H\"{o}lder's Inequality. By continuity of the exponential function, we have
\begin{equation}
    \lim_{\theta \rightarrow 0} \bar{f}(\omega,\theta) = \lim_{\theta \rightarrow 0} \textstyle Z'(\omega) \exp(\frac{-\theta}{2}Z'(\omega)) = Z'(\omega) \;\;\; \forall \omega \in \Omega.
\end{equation}
As we have verified the conditions that are required for \cite[Thm. 2.27a]{folland1999real}, we apply this result to interchange the limit and the integral and thereby conclude \eqref{my22}. The derivation of \eqref{my33} uses a similar argument.\\\\
The proof of 
\begin{equation}
    \underset{\theta \rightarrow 0^+}{\lim} \textstyle\frac{-2}{\theta}\log E_x^\pi(\exp(\frac{-\theta}{2} Z')) = \underset{\theta \rightarrow 0^-}{\lim} \textstyle\frac{-2}{\theta}\log E_x^\pi(\exp(\frac{-\theta}{2} Z')) = E_x^\pi(Z')
\end{equation}
follows from $E_x^\pi(\exp(\frac{-\theta}{2}Z'))$ being positive and finite for all $\theta \in [a,b]$, \eqref{my1}--\eqref{my33}, and L'H\^{o}pital's Rule. \\\\Remark: If $c$ and $c_N$ are bounded, then $Z'$ is an element of $L^2$, in particular. $L^p$ spaces are formally presented by \cite[Chap. 6]{folland1999real}, for example.

\subsection{Note Regarding Mean-Variance Approximation}
Here, we provide details regarding Eq. (6) from the main paper. Let $Y$ be a non-negative random variable on a probability space $(\Omega, \mathcal{F}, \mu)$. Let $E(g(Y)) := \int_\Omega g(Y) \mathrm{d}\mu$ denote the expectation of $g(Y)$, where $g : \mathbb{R} \rightarrow \mathbb{R}$ is a Borel-measurable function. We paraphrase the statement of interest from the main paper: if the magnitude of $\theta$ is sufficiently small and if there is an $M < +\infty$ such that $E(Y^n) \leq M$ for all $n \in \mathbb{N}$, then the EU of $Y$ approximates a weighted sum of the expectation $E(Y)$ and variance $\text{var}(Y)$,
\begin{equation*}
    \rho_\theta(Y) := \textstyle\frac{-2}{\theta}\log E(\exp(\frac{-\theta}{2} Y)\big) \approx E(Y) - \textstyle\frac{\theta}{4}\text{var}(Y).
\end{equation*}
By the definition of the exponential function, e.g., see \cite[Eq. 1, p. 1]{rudin2006real}, it holds that
\begin{equation}\label{expdef}
    \exp({\textstyle \frac{-\theta}{2}y}) = \sum_{n =0}^\infty \frac{({\textstyle\frac{-\theta}{2}y})^n}{n!}
\end{equation}
for all $y \in \mathbb{R}$. Recall that we consider $\theta \in \Theta \subseteq (-\infty,0)$, and thus,
\begin{equation}
    h_n := \frac{({\textstyle\frac{-\theta}{2}Y})^n}{n!}
\end{equation}
is a non-negative Borel-measurable function for each $n \in \mathbb{N}$. Since any series of non-negative Borel-measurable functions can be integrated term by term \cite[Corollary 1.6.4 (a), p. 46]{ash1972probability}, it holds that
%= \sum_{n = 0}^\infty E\left(\frac{({\textstyle\frac{-\theta}{2}Y})^n}{n!}\right)
\begin{equation}\label{phitheta}
    E(\exp({\textstyle \frac{-\theta}{2}Y}))  = \sum_{n = 0}^\infty \frac{({\textstyle\textstyle\frac{-\theta}{2}})^n}{n!} E(Y^n) = 1 + \underbrace{{\textstyle\textstyle\frac{-\theta}{2}} E(Y) + \frac{({\textstyle\textstyle\frac{-\theta}{2}})^2}{2} E(Y^2) + \sum_{n = 3}^\infty \frac{({\textstyle\textstyle\frac{-\theta}{2}})^n}{n!} E(Y^n)}_{\phi_\theta},
\end{equation}
where each integral is guaranteed to exist (i.e., not be of the form $+\infty - \infty$) because each function inside each integral is non-negative and Borel measurable. \\\\Now, recall the following relation for the natural logarithm,
\begin{equation}\label{my4}
    \log(1 + z) = z - \frac{z^2}{2} + \frac{z^3}{3} - \frac{z^4}{4} + \dots \;\;\; \text{for }-1 <z \leq 1,
\end{equation}
e.g., see \cite[Example 2, pp. 212--213]{ross2013elementary}. Since $\frac{-\theta}{2}Y$ is non-negative and the exponential is increasing, it holds that $\exp(\frac{-\theta}{2}Y) \geq \exp(0) = 1$ everywhere, and thus,  $E(\exp({\textstyle \frac{-\theta}{2}Y})) \geq 1$. In addition, we use \eqref{phitheta} and the assumed existence of an $M < +\infty$ such that $E(Y^n) \leq M$ for all $n \in \mathbb{N}$ to find that
\begin{equation}\begin{aligned}\label{my5}
   0 \leq E(\exp({\textstyle \frac{-\theta}{2}Y})) - 1 = \phi_\theta & = {\textstyle\textstyle\frac{-\theta}{2}} E(Y) + \frac{({\textstyle\textstyle\frac{-\theta}{2}})^2}{2} E(Y^2) + \sum_{n = 3}^\infty \frac{({\textstyle\textstyle\frac{-\theta}{2}})^n}{n!} E(Y^n)\\
   & \leq {\textstyle\textstyle\frac{-\theta}{2}} M + \frac{({\textstyle\textstyle\frac{-\theta}{2}})^2}{2} M + \sum_{n = 3}^\infty \frac{({\textstyle\textstyle\frac{-\theta}{2}})^n}{n!} M\\
   & = M \sum_{n = 1}^\infty \frac{({\textstyle\textstyle\frac{-\theta}{2}})^n}{n!}.
\end{aligned}\end{equation}
By the definition of the exponential, e.g., use \eqref{expdef} with $y=1$, it holds that
\begin{equation}\label{my6}
   \exp({\textstyle \frac{-\theta}{2}}) = \sum_{n =0}^\infty \frac{({\textstyle\frac{-\theta}{2}})^n}{n!} = 1 + \sum_{n =1}^\infty \frac{({\textstyle\frac{-\theta}{2}})^n}{n!}, 
\end{equation}
and by \eqref{my5} and \eqref{my6},
\begin{equation}\label{my7}
    0 \leq \phi_\theta \leq M (\exp({\textstyle \frac{-\theta}{2}}) - 1).
\end{equation}
Note that there is a $\theta < 0$ whose magnitude is sufficiently small so that $0 \leq \phi_\theta \leq 1$ holds. Using such a $\theta$, we apply $E(\exp({\textstyle \frac{-\theta}{2}Y})) = 1 + \phi_\theta$ \eqref{phitheta} and \eqref{my4} with $z = \phi_\theta$ to write
\begin{equation}
    \log E(\exp({\textstyle \frac{-\theta}{2}Y})) \overset{\eqref{phitheta}}{=} \log(1 + \phi_\theta) \overset{\eqref{my4}}{=} \phi_\theta - \frac{\phi_\theta^2}{2} + \frac{\phi_\theta^3}{3} - \frac{\phi_\theta^4}{4} + \dots.
\end{equation}
By discarding the terms of order three or greater, we have the following approximation,
\begin{equation}
    \log E(\exp({\textstyle \frac{-\theta}{2}Y})) \approx \phi_\theta - \frac{\phi_\theta^2}{2},
\end{equation}
whose accuracy improves when we have chosen $\theta$ so that $\phi_\theta$ is closer to zero.
By substituting the expression for $\phi_\theta$, see \eqref{phitheta}, and discarding terms of order three or greater, we have
\begin{equation}\begin{aligned}
    \log E(\exp({\textstyle \frac{-\theta}{2}Y})) & \approx {\textstyle\textstyle\frac{-\theta}{2}} E(Y) + \frac{({\textstyle\textstyle\frac{-\theta}{2}})^2}{2} E(Y^2) -  \frac{({\textstyle\textstyle\frac{-\theta}{2}} E(Y))^2}{2} \\ & = {\textstyle\textstyle\frac{-\theta}{2}} E(Y) + \frac{({\textstyle\textstyle\frac{-\theta}{2}})^2}{2}\text{var}(Y).
\end{aligned}\end{equation}
Finally, by multiplying by $\frac{-2}{\theta}$, we obtain the desired approximation,
\begin{equation}
    { \textstyle \frac{-2}{\theta} } \log E(\exp({\textstyle \frac{-\theta}{2}Y})) \approx E(Y) - \textstyle \frac{\theta}{4}\text{var}(Y).
\end{equation}

\section{Some Details about CVaR Optimal Control}
For convenience, we first repeat some information from the main paper. The function $J^* : S \times \mathbb{R} \rightarrow \mathbb{R}$ is defined by
\begin{equation*}
     J^*(x,s) := \inf_{\pi \in \Pi} E_x^\pi(\max\{Z' - s,0 \}),
\end{equation*}
where $Z'$ is a non-negative, everywhere-bounded cumulative random cost incurred by a control system over time. In particular, each realization of $Z'$ is an element of $[0, \bar{a}]$, where $\bar{a} \in \mathbb{R}_+$. Details regarding the precise meaning of $E_x^\pi(\cdot)$ in the definition of $J^*$ will be provided below.\\\\
We define $\mathcal{Z}:= [-\bar{a},\bar{a}] \subseteq \mathbb{R}$. \\\\
$\Pi$ is a class of policies that are history-dependent through the augmented state $(X_t,S_t)$. Any $\pi \in \Pi$ takes the form $\pi = (\pi_0,\pi_1,\dots,\pi_{N-1})$, where $\pi_t(\cdot|\cdot,\cdot)$ is a Borel-measurable stochastic kernel on $A$ given $S \times \mathcal{Z}$ for each $t$.\\\\
In the main paper, we have defined $\Omega := (S \times A)^N \times S$, and we have stated that $P_x^\pi$ is a probability measure on $(\Omega,\mathcal{B}_\Omega)$ that is parametrized by an initial condition $x \in S$ and a policy $\pi$. We have said that the notation $E_x^\pi(\cdot)$ denotes the expectation with respect to $P_x^\pi$. Now, in the case of CVaR, we use different definitions for $\Omega$ and $P_{x}^\pi$ to accommodate an extended state space. In particular, we use $\Omega := (S \times \mathcal{Z} \times A)^N \times S \times \mathcal{Z}$. Let $\delta_y$ denote the Dirac measure on $(\mathcal{M},\mathcal{B}_\mathcal{M})$ concentrated at $y \in \mathcal{M}$, where $\mathcal{M}$ is a metrizable space. Let $Q(\cdot|\cdot,\cdot)$ be the \emph{transition law}, which is a Borel-measurable stochastic kernel on $S$ given $S \times A$. That is, if $(x_t,u_t) \in S \times A$ is the realization of $(X_t,U_t)$, then the probability that $X_{t+1}$ is in $B \in \mathcal{B}_S$ is given by
\begin{equation}
    Q(B|x_t,u_t) := p(\{w_t \in D : f(x_t,u_t,w_t) \in B \}|x_t,u_t).
\end{equation}
Let $(x,s) \in S \times \mathcal{Z}$ and $\pi \in \Pi$ be given. $P_{x}^\pi$ takes the following form on measurable rectangles in $\Omega$,
\begin{equation}\label{keyprobabilitymeasure}\begin{aligned}
    & P_{x}^\pi(\underline{S}_0 \times \underline{\mathcal{Z}}_0 \times \underline{A}_0 \times \underline{S}_1 \times \underline{\mathcal{Z}}_1 \times \underline{A}_1 \times \cdots \times \underline{S}_{N-1} \times \underline{\mathcal{Z}}_{N-1} \times \underline{A}_{N-1} \times \underline{S}_{N} \times \underline{Z}_{N}) = \\
    & \textstyle \int_{\underline{S}_0} \int_{\underline{\mathcal{Z}}_0} \int_{\underline{A}_0} \int_{\underline{S}_1} \int_{\underline{\mathcal{Z}}_1} \int_{\underline{A}_1} \cdots \int_{\underline{S}_{N-1}} \int_{\underline{\mathcal{Z}}_{N-1}} \int_{\underline{A}_{N-1}} \int_{\underline{S}_{N}}  \int_{\underline{\mathcal{Z}}_{N}} \delta_{(s_{N-1} - c'(x_{N-1},u_{N-1}))}(\mathrm{d}s_N) \; Q(\mathrm{d}x_N|x_{N-1},u_{N-1}) \\ & \hphantom{==} \pi_{N-1}(\mathrm{d}u_{N-1}|x_{N-1},s_{N-1}) \; \delta_{(s_{N-2} - c'(x_{N-2},u_{N-2}))}(\mathrm{d}s_{N-1}) \;  Q(\mathrm{d}x_{N-1}|x_{N-2},u_{N-2})  \cdots  \\ & \hphantom{==} \pi_{1}(\mathrm{d}u_1|x_1,s_1) \; \delta_{(s_0 - c'(x_0,u_0))}(\mathrm{d}s_1) \;  Q(\mathrm{d}x_1|x_0,u_0) \; \pi_{0}(\mathrm{d}u_0|x_0,s_0) \; \delta_s(\mathrm{d}s_0) \; \delta_x(\mathrm{d}x_0),
\end{aligned}\end{equation}
where $\underline{S}_t \in \mathcal{B}_S$, $\underline{\mathcal{Z}}_t \in \mathcal{B}_\mathcal{Z}$, and $\underline{A}_t \in \mathcal{B}_A$ for each $t$ \cite[Prop. 7.28, pp. 140--141]{bertsekas2004stochastic}.
Note that $P_{x}^\pi$ depends on $s$, which we do not write explicitly to follow the convention in the literature, e.g., see \cite{bauerleott}.

\end{document}